%% file: main.tex
\newif\ifneuripsformat
\neuripsformatfalse

\ifneuripsformat
\PassOptionsToPackage{authoryear,sort&compress,round}{natbib}
\documentclass{article}
\else
\documentclass[11pt,a4paper]{swe_evo_format}
\fi

\ifneuripsformat
\usepackage[eandd]{neurips_2026}
\else
\usepackage[authoryear,sort&compress,round]{natbib}
\fi

\usepackage[utf8]{inputenc}
\usepackage[T1]{fontenc}
\usepackage{hyperref}
\usepackage{url}
\usepackage{booktabs}
\usepackage{amsfonts}
\usepackage{nicefrac}
\usepackage{microtype}
\usepackage{xcolor}

\usepackage{latexsym}
\usepackage{amsmath}
\usepackage{amssymb}
\usepackage{amsthm}
\usepackage{graphicx}
\usepackage{xspace}
\usepackage{subcaption}
\usepackage{multirow}
\usepackage[export]{adjustbox}
\usepackage{tikz}
\usepackage{pifont}
\usepackage{xurl}
\usepackage{enumitem}
\usepackage[hang,flushmargin]{footmisc}
\usepackage{caption}
\usepackage{listings}
\usepackage{pdfpages}
\usepackage{algorithm}
\usepackage{algorithmic}
\usepackage[most]{tcolorbox}
\ifneuripsformat\else
\usepackage{fontawesome5}
\fi

\definecolor{sweevoBlue}{RGB}{0, 90, 156}
\definecolor{sweevoTeal}{RGB}{0, 128, 128}
\definecolor{sweevoGray}{RGB}{85, 85, 85}

\newtcolorbox{problemstatementboxA}[1][]{%
  enhanced jigsaw,
  width=\linewidth,
  sharp corners=south,
  colback=gray!10,
  colframe=blue,
  coltitle=white,
  colbacktitle=blue,
  fonttitle=\bfseries,
  fontupper=\scriptsize,
  title=Problem Statement: iterative\_\_dvc\_0.33.1\_0.34.0,
  boxrule=0.7pt, leftrule=0.7pt, rightrule=0.7pt,
  bottomrule=0.7pt, toprule=0.7pt, titlerule=0pt,
  arc=2mm,
  top=1.0ex, bottom=1.0ex, left=1.0em, right=1.0em,
  before skip=1ex plus 0.5ex minus 0.2ex,
  after skip=1ex plus 0.5ex minus 0.2ex,
  before upper={\let\href\problemstatementhref\sloppy\setlength{\emergencystretch}{3em}},
  #1
}

\newtcolorbox{problemstatementboxB}[1][]{%
  enhanced jigsaw,
  width=\linewidth,
  sharp corners=south,
  colback=gray!10,
  colframe=blue,
  coltitle=white,
  colbacktitle=blue,
  fonttitle=\bfseries,
  fontupper=\scriptsize,
  title=Problem Statement: dask\_\_dask\_2023.6.1\_2023.7.0,
  boxrule=0.7pt, leftrule=0.7pt, rightrule=0.7pt,
  bottomrule=0.7pt, toprule=0.7pt, titlerule=0pt,
  arc=2mm,
  top=1.0ex, bottom=1.0ex, left=1.0em, right=1.0em,
  before skip=1ex plus 0.5ex minus 0.2ex,
  after skip=1ex plus 0.5ex minus 0.2ex,
  before upper={\let\href\problemstatementhref\sloppy\setlength{\emergencystretch}{3em}},
  #1
}

\newtcolorbox{keyfindings}{
  enhanced,
  colback=sweevoBlue!3,
  colframe=sweevoBlue!60,
  boxrule=0.5pt,
  left=8pt, right=8pt, top=6pt, bottom=6pt,
  arc=2pt,
  fonttitle=\bfseries\color{sweevoBlue},
  title=Key Findings,
  attach boxed title to top left={yshift=-2mm, xshift=5mm},
  boxed title style={
    colback=white,
    colframe=white,
    boxrule=0pt
  }
}

\newcommand{\tool}{\textsc{SWE-EVO}\xspace}

\newcommand{\problemstatementhref}[2]{\textcolor{cyan}{#2}}

\renewcommand{\cite}{\citep}

\AtBeginDocument{%
  }

\lstdefinelanguage{Python}{
    morekeywords={class, def, return, try, except, raise, from},
    keywordstyle=\color{blue},
    stringstyle=\color{green},
    commentstyle=\color{gray},
    morecomment=[l]{\#},
}
\lstdefinelanguage{diff}{
    morecomment=[f][\color{green}]{+},
    morecomment=[f][\color{red}]{-},
    morecomment=[f][\color{blue}]{@@},
}
\lstdefinelanguage{errorlog}{
    morecomment=[f][\color{red}]{E},
    morecomment=[f][\color{magenta}]{?},
}

\ifneuripsformat
\title{SWE-EVO: Benchmarking Coding Agents in Long-Horizon Software Evolution Scenarios}
\author{%
  Tue Le\thanks{Equal contribution.} \\
  FPT Software AI Center \\
  \texttt{tueldt1@fpt.com} \\
  \And
  Minh Vu Thai Pham\footnotemark[1] \\
  FPT Software AI Center \\
  \texttt{minhpvt@fpt.com} \\
  \And
  Dung Nguyen Manh \\
  School of Computing and Information Systems \\
  University of Melbourne \\
  \texttt{manhdung.nguyen.1@student.unimelb.edu.au} \\
  \And
  Huy Nhat Phan \\
  FPT Software AI Center \\
  \texttt{huypn168@gmail.com} \\
  \And
  Nghi D. Q. Bui\thanks{Project lead.} \\
  Center of AI Research, VinUniversity Ha Noi, Viet Nam \\
  \texttt{bdqnghi@gmail.com}
}
\else
\title{\textcolor{blue!70!black}{\textsc{SWE-EVO}}: Benchmarking Coding Agents in Long-Horizon Software Evolution Scenarios}

\author{%
  Tue Le\textsuperscript{1,*} \quad
  Minh Vu Thai Pham\textsuperscript{1,*} \quad
  Dung Nguyen Manh\textsuperscript{2} \quad
  Huy Nhat Phan\textsuperscript{1} \quad
  Nghi D. Q. Bui\textsuperscript{3,\dag}\\
  \textsuperscript{1}FPT Software AI Center, Ha Noi, Viet Nam\\
  \textsuperscript{2}School of Computing and Information Systems, University of Melbourne\\
  \textsuperscript{3}Center of AI Research, VinUniversity, Ha Noi, Viet Nam
}

\githuburl{https://github.com/SWE-EVO/SWE-EVO}
\fi

\begin{document}

\ifneuripsformat
\maketitle

\begin{abstract}
Existing benchmarks for AI coding agents focus on isolated, single-issue tasks such as fixing a bug or adding a small feature. However, real-world software engineering is a long-horizon endeavor: developers interpret high-level requirements, coordinate changes across many files, and evolve codebases over multiple iterations while preserving functionality. We introduce \tool, a benchmark for this long-horizon software evolution challenge. Constructed from release transitions in seven mature open-source Python projects, \tool comprises 48 release-sized tasks requiring multi-step modifications spanning an average of 21 files, validated against test suites averaging 874 tests per instance. Experiments reveal a striking capability gap: the best model reaches only 25\% on \tool, while \texttt{gpt-5.2} drops from 72.80\% on SWE-Bench Verified to 22.92\% on \tool, showing that current agents struggle with sustained, multi-file reasoning. We also propose \emph{Fix Rate}, a metric capturing partial progress on these complex, long-horizon tasks.
\end{abstract}
\else
\begin{abstract}
Existing benchmarks for AI coding agents focus on isolated, single-issue tasks such as fixing a bug or adding a small feature. However, real-world software engineering is a long-horizon endeavor: developers interpret high-level requirements, coordinate changes across many files, and evolve codebases over multiple iterations while preserving functionality. We introduce \tool, a benchmark for this long-horizon software evolution challenge. Constructed from release transitions in seven mature open-source Python projects, \tool comprises 48 release-sized tasks requiring multi-step modifications spanning an average of 21 files, validated against test suites averaging 874 tests per instance. Experiments reveal a striking capability gap: the best model reaches only 25\% on \tool, while \texttt{gpt-5.2} drops from 72.80\% on SWE-Bench Verified to 22.92\% on \tool, showing that current agents struggle with sustained, multi-file reasoning. We also propose \emph{Fix Rate}, a metric capturing partial progress on these complex, long-horizon tasks.
\end{abstract}

\renewcommand{\thefootnote}{}
\footnotetext{\textsuperscript{*}Equal contribution \hspace{2em} \textsuperscript{\dag}Project lead\\
Correspondence to: Tue Le \texttt{<tueldt1@fpt.com>}, Minh Vu Thai Pham \texttt{<minhpvt@fpt.com>},\\
Dung Nguyen Manh \texttt{<manhdung.nguyen.1@student.unimelb.edu.au>}, Huy Nhat Phan \texttt{<huypn168@gmail.com>}, Nghi D. Q. Bui \texttt{<bdqnghi@gmail.com>}.}
\renewcommand{\thefootnote}{\arabic{footnote}}
\maketitle
\fi

\input{sections/intro}
\input{sections/related}
\input{sections/dataset}
\input{sections/experiments}
\input{sections/conclusion}

\newpage
\ifneuripsformat
\bibliographystyle{abbrvnat}
\else
\bibliographystyle{plainnat}
\fi
\bibliography{custom}

\newpage
\appendix
\input{sections/appendix}

\ifneuripsformat
\clearpage
\input{checklist}
\fi

\end{document}

%% file: sections/intro.tex
\section{Introduction}
Large language models (LLMs) have achieved remarkable progress in automating software engineering (SE) tasks, including code generation~\cite{wei2023magicoder, chen2021evaluating, wang2023codet5+, li2022competition, zhuo2024bigcodebench, bui2023codetf, manh2023vault, to2023functional}, bug fixing~\cite{jimenez2023swe, xia2024agentless}, and test synthesis~\cite{chen2022codet, wang2024testeval, jain2024testgeneval}. These advancements have facilitated the emergence of AI-powered coding agents capable of assisting or automating key aspects of the software development lifecycle~\cite{zhang2023survey, fan2023large, gao2025current, he2025llm}.

Building on these capabilities, multi-agent systems have rapidly emerged to address long-horizon challenges in software engineering. By coordinating specialized agents for navigation, localization, patching, and verification, these frameworks improve scalability and performance over single-agent architectures. Recent systems~\cite{yang2024swe,wang2024openhands,phan2024hyperagent,nguyen2025agilecoder} illustrate this shift in real-world repositories. Industry adoption reflects this momentum: over 90\% of engineering teams now integrate generative AI into SE workflows, up from 61\% in 2024~\citep{dora2025ai}.


To evaluate these agents rigorously, benchmarks have become essential. Early efforts like HumanEval~\cite{humaneval} focused on function-level code completion~\cite{chen2021evaluating}, while SWE-Bench~\cite{jimenez2023swe} curated real-world GitHub issues requiring verifiable patches for isolated problems~\cite{jimenez2023swe}. SWE-Bench is now a de facto standard for assessing multi-agent capabilities in practical coding scenarios.

\begin{figure*}[t]
        \centering
        \includegraphics[width=0.9\textwidth]{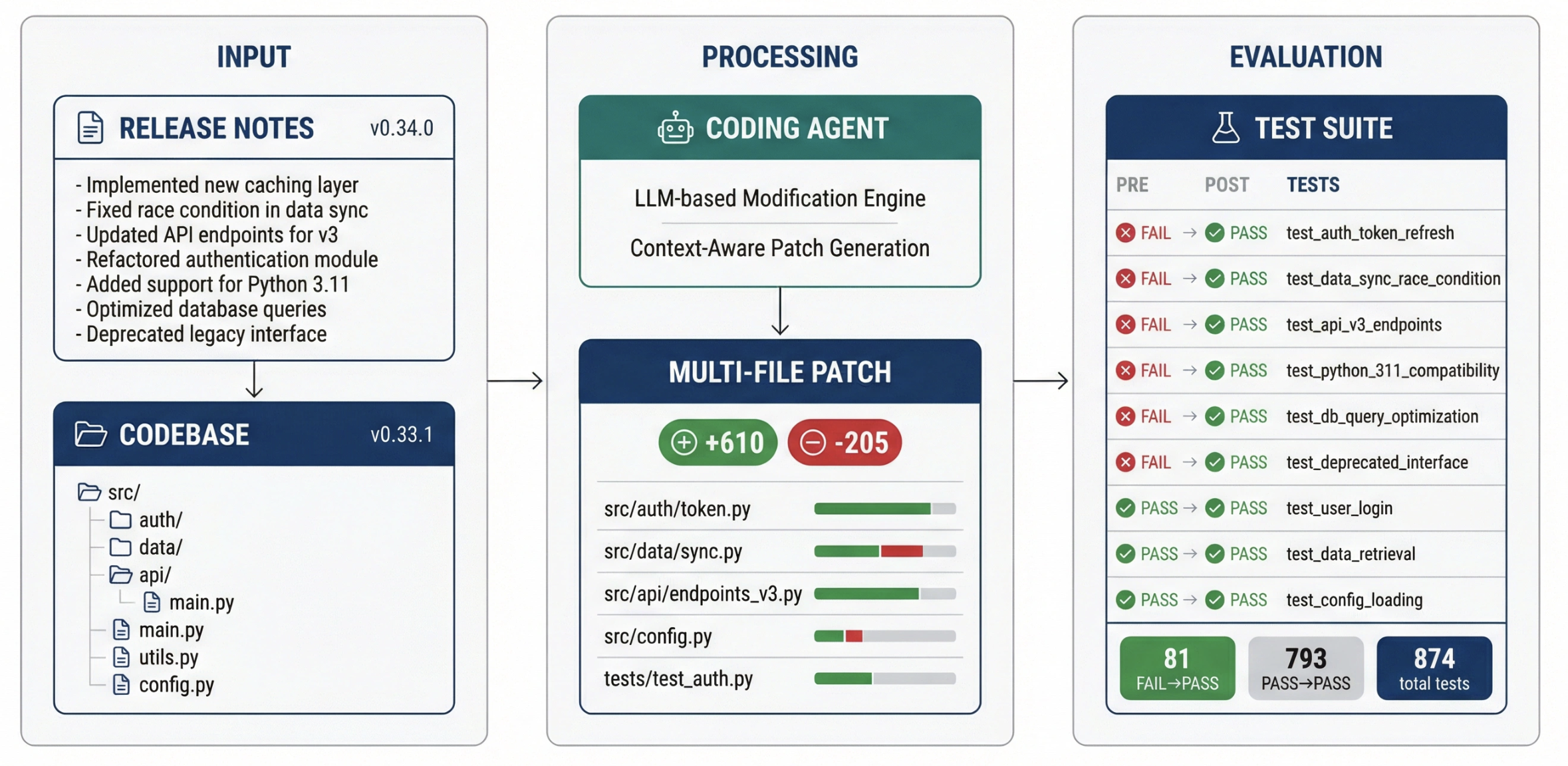}
        \caption{Overview of the \tool task pipeline. \textbf{Input}: the agent receives release notes specifying multiple changes (bug fixes, new features, maintenance) alongside the full codebase at the start version. \textbf{Processing}: a coding agent generates a multi-file patch spanning dozens of files. \textbf{Evaluation}: the patch is validated against a comprehensive test suite comprising FAIL$\to$PASS tests (verifying the required changes) and PASS$\to$PASS tests (ensuring no regressions).}
        \label{fig:swe-evo-pipeline}
\end{figure*}

While recent models reach around 75\% on SWE-Bench-Verified (e.g., \texttt{GPT-5.2}~\citep{openai_gpt52_2025}) and around 40\% on the full leaderboard (e.g., OpenCSG Starship at 39.67\%~\citep{jimenez2024swebench}), the benchmark shows signs of saturation on isolated tasks. More importantly, SWE-Bench focuses on discrete issue resolution and does not capture the core SE challenge: continuous evolution of existing systems~\citep{kaur2015review, singh2019analysis}. In practice, up to 80\% of effort is spent maintaining and evolving legacy code, requiring coordinated changes across modules, versions, and specifications~\citep{kaur2015review, singh2019analysis}.


This gap between benchmark tasks and real-world evolution scenarios motivates our central research question: \textit{Given an existing codebase, can multi-agent LLM systems autonomously evolve the system in response to dynamic input requirements, demonstrating sustained planning, adaptability, and innovation across long-horizon tasks?}

To address this gap, we introduce \tool, a benchmark for autonomous software evolution rather than single-issue repair. \tool uses release notes, commit histories, and versioned snapshots from mature open-source Python projects to construct tasks where agents interpret Software Requirement Specifications (SRS), plan multi-step modifications, and evolve codebases across releases. \tool comprises 48 release-sized tasks across 7 repositories (Table~\ref{tab:swe_stats} and Figure~\ref{fig:repo_dist}). We describe construction in Section~\ref{sec:dataset} and results in Section~\ref{sec:experiments}.

\begin{figure*}[t]
        \centering
        \includegraphics[width=0.9\textwidth]{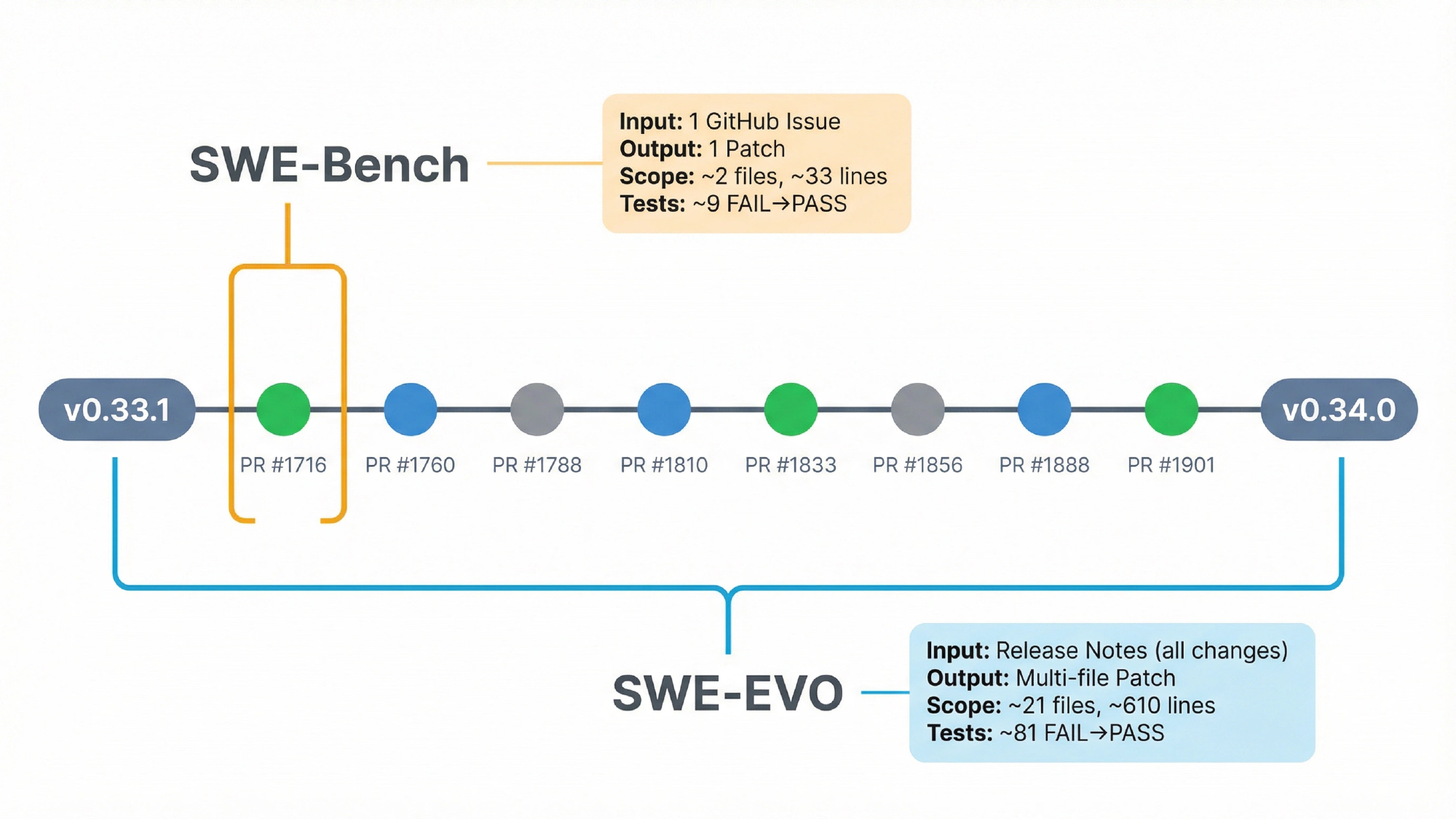}
        \caption{Comparison of SWE-Bench and \tool. SWE-Bench tasks agents with resolving a single GitHub issue to produce one patch. \tool requires agents to interpret release notes and implement comprehensive changes that resolve multiple PRs, develop new features, and fix multiple bugs to evolve the codebase to a new version.}
        \label{fig:comparison}
\end{figure*}

Figure~\ref{fig:swe-evo-pipeline} illustrates the \tool pipeline: from release notes and a start-version codebase, the agent generates a multi-file patch validated by comprehensive tests. Software evolution (Figure~\ref{fig:software-evolution} in Appendix~\ref{sec:evolution_model_appendix}) requires SRS alignment, holistic program understanding, and coordinated changes, unlike SWE-Bench's isolated issue setting (Figure~\ref{fig:comparison}).

Our evaluation with OpenHands, SWE-agent, and 18 models reveals a significant capability gap: the best model (\texttt{gpt-5.4}) resolves only 25\% of \tool, while \texttt{gpt-5.2} drops from 72.80\% on SWE-Bench Verified to 22.92\%. Larger models generally outperform smaller variants. Failure analysis suggests that stronger models more often misinterpret nuanced release notes, while weaker models struggle with tool use and syntax errors.

%% file: sections/related.tex
\section{Related Work}

\subsection{Code Generation and SE Benchmarks}

Code generation evaluation has moved from function-level tasks~\citep{chen2021evaluating,austin2021program,hendrycks2021measuring} to repository-level SE challenges. Early benchmarks such as HumanEval, MBPP, and APPS remain largely single-file and increasingly saturated~\citep{zhou2023language}, while newer efforts such as CodeMMLU~\citep{nguyencodemmlu} and CodeWiki~\citep{hoang2025codewiki} broaden evaluation to code understanding and documentation. SWE-bench~\citep{jimenez2024swebench} shifted the field to verifiable patches for real GitHub issues, with curated subsets~\citep{sbv} and extensions across languages~\citep{zan2025multi}, multimodal inputs~\citep{yang2024swe2}, live collection~\citep{zhang2025swe,badertdinov2025swe}, and enterprise-scale settings~\citep{deng2025swe}. EnConda-Bench~\citep{kuang2025enconda} isolates environment-configuration trajectories, including setup planning, diagnosis, repair, and execution. \tool instead targets version-to-version software evolution, where agents must implement release-level changes in the codebase rather than fix injected configuration faults. Prior benchmarks therefore emphasize isolated issue resolution or narrow subskills more than multi-step evolution across commits and longer planning horizons.

\subsection{Software Engineering Agents}

Interactive bug-fixing work~\citep{xia2023conversational,xia2024automated,chen2023teaching,yuan2024evaluating} and autonomous systems such as Devin AI~\citep{devin} established the basis for SE agents. SWE-agent~\citep{yang2024swe} emphasized the agent--computer interface; OpenHands~\citep{wang2024openhands} presented a multi-agent platform evaluated on 15 benchmarks; AutoCodeRover~\citep{zhang2024autocoderover} paired LLMs with AST-based search; and Agentless~\citep{xia2024agentless} showed that a simple localization-repair pipeline can remain competitive. Multi-agent designs further explore role specialization and coordination, including AgentCoder~\citep{huang2023agentcoder}, AgileCoder~\citep{nguyen2025agilecoder}, and CodeAct~\citep{wang2024executable}. Model-side progress includes post-training on real SE data~\citep{wei2025swerl,luodeepswe,deepseekv31,chen2025minimax,team2025kimi,carbonneaux2025cwm} and synthetic-data generation~\citep{pham2025swesynth}. A related line studies self-evolving agents and tool construction~\citep{robeyns2025self,zhang2025darwin,wang2025huxley,llmastool,wang2024voyager,qian2024creatortoolcreationdisentangling,wang2024troveinducingverifiableefficient,qiu2025alitageneralistagentenabling} (Appendix~\ref{sec:self_evolving_appendix}).

\subsection{Context Engineering for Long-Horizon Agents}

Long-horizon agents depend on effective context management, which is central to \tool because tasks unfold across large codebases and many turns. Surveys~\citep{mei2025survey,hua2025context} frame the area around retrieval, processing, and management, tracing a path from prompt engineering and RAG to context engineering. Representative methods include compression~\citep{ge2023context}, retrieval schemes such as Self-RAG~\citep{asai2023self}, RAPTOR~\citep{sarthi2024raptor}, and GraphRAG~\citep{gutierrez2024hipporag}, and memory systems such as MemGPT~\citep{packer2023memgpt} and hierarchical storage~\citep{zhong2023memorybank} that extend beyond single sessions. Meta Context Engineering~\citep{ye2026meta} pushes this further by optimizing context assembly, reporting 89.1\% on SWE-bench Verified versus 70.7\% for hand-engineered baselines. These ideas are directly relevant to \tool, where agents must reason over long specifications and patches spanning many files.

%% file: sections/dataset.tex
\section{{\tool} Dataset}
\label{sec:dataset}

\tool is a benchmark constructed from release notes and version histories of popular open-source repositories, capturing real software evolution scenarios where coding agents must interpret high-level software requirement specifications, plan and implement multi-step modifications across versions, and ensure that the evolved system passes validation tests aligned with those specifications.

\begin{figure*}[ht]
\centering

\begin{minipage}[t]{0.54\textwidth}
\vspace{0pt}%
\centering
\small
\begin{tabular}{l l rr}
\toprule
\cmidrule(lr){3-4}
 & & \textbf{Mean} & \textbf{Max} \\
\midrule
\multirow{1}{*}{Issue Text} & Length (Words) & 2390.5 & 22344 \\
\midrule
\multirow{2}{*}{Codebase}
 & \# Files (non-test) & 363 & 1046 \\
 & \# Lines (non-test) & 78K & 272K \\
\midrule
\multirow{3}{*}{Gold Patch}
 & \# Lines edited & 610.5 & 4113 \\
 & \# Files edited & 20.9 & 105 \\
 & \# Func. edited & 51.0 & 379 \\
\midrule
\multirow{2}{*}{Tests}
 & \# Fail to Pass & 81.4 & 2774 \\
 & \# Total & 874.0 & 8552 \\
\bottomrule
\end{tabular}
\captionsetup{type=table}
\captionof{table}{Average and maximum numbers characterizing different attributes of a \tool\ task instance. Statistics are micro-averages calculated without grouping by repository.}
\label{tab:swe_stats}
\end{minipage}\hfill
\begin{minipage}[t]{0.40\textwidth}
\vspace{0pt}%
\centering
\includegraphics[valign=t,width=\linewidth]{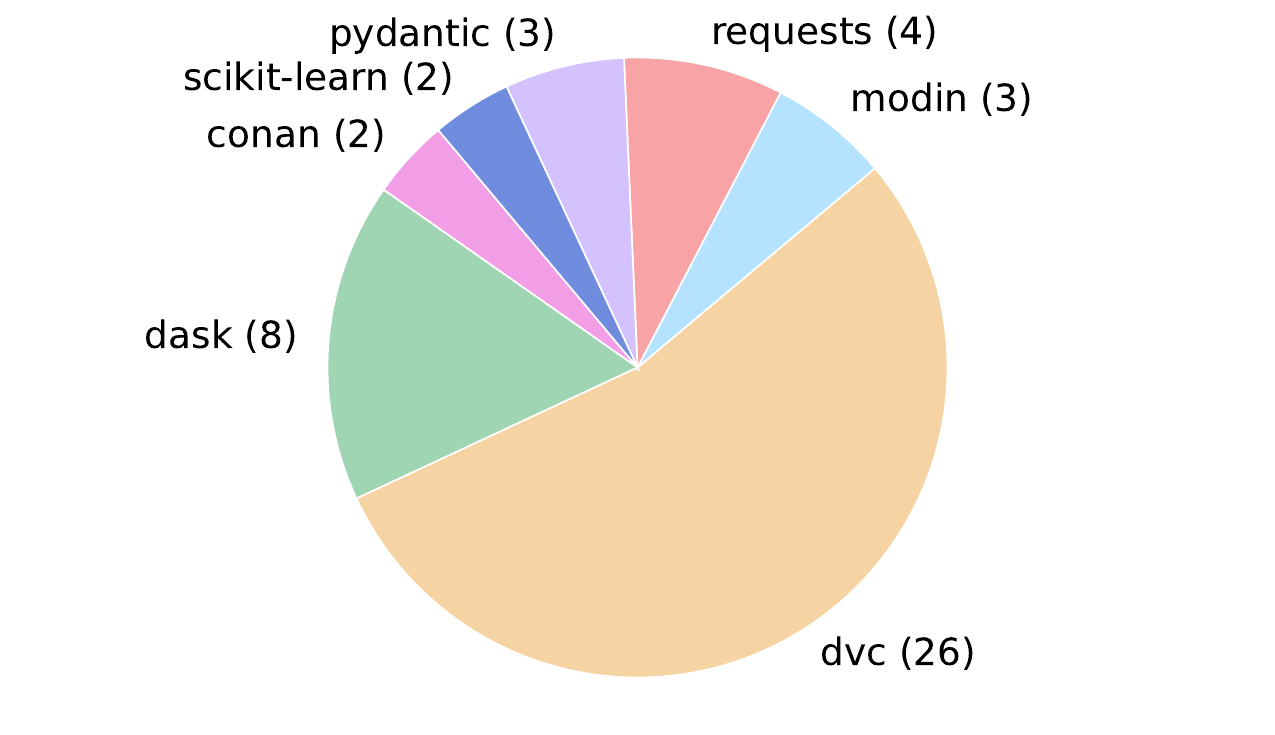}
\captionof{figure}{Distribution of \tool\ tasks (in parenthesis) across 7 open source GitHub repositories that each contains the source code for a popular, widely downloaded PyPI package.}
\label{fig:repo_dist}
\end{minipage}

\end{figure*}

\subsection{Benchmark Construction}

The construction of \tool consists of three major phases: (1) repository selection and data scraping, (2) candidate selection and filtering, and (3) execution-based filtering. We describe each phase in turn.

\textbf{Stage I: Repository Selection and Data Scraping.} 
To maximize reproducibility and comparability, we inherit repositories and execution environments from SWE-bench and it keeps our benchmark ``plug-and-play'' for existing SWE agents. Concretely, we begin by collecting samples from SWE-bench~\citep{jimenez2024swebench} and from SWE-gym~\citep{pan2024training} as our seed pool of task instances. Both resources already give us: a real repository snapshot, an executable environment, and a set of tests tied to a human-authored change. 


\textbf{Stage II: Candidate Selection and Filtering.} Unlike SWE-Bench, which frames tasks as resolving a single issue, we target evolving a codebase between two release versions. We therefore create candidate instances by selecting only those samples whose base commit corresponds exactly to a version tag of the repository (i.e., a release snapshot). For each such candidate, we define the problem statement as the release-note delta between that version and its next tagged version; the agent's job is to implement the specified changes across the codebase. This framing treats a release as the unit of evolution, matching how users and maintainers typically receive a mixture of fixes, features, and maintenance changes in open-source projects. Importantly, \tool is not intended to imitate a single clean pull request or one atomic commit. Instead, it evaluates whether an agent can coordinate the full release-level behavioral delta. This formulation intentionally withholds an oracle decomposition into a sequence of separately specified PRs: agents may still work iteratively during their trajectory, but evaluation is based on the final evolved repository state. A sequential PR-by-PR variant is a useful future setting, but it would provide cleaner intermediate supervision and move closer to repeated single-issue repair.

\textbf{Stage III: Execution-Based Filtering.} Following SWE-Bench, we validate each candidate by applying the instance's test patch content and logging test outcomes \emph{before} and \emph{after} the remaining patch content is applied. We retain only instances that exhibit at least one \texttt{FAIL\_TO\_PASS} test (i.e., a test that fails pre-patch and passes post-patch), ensuring a measurable behavioral change attributable to the required evolution. We additionally discard candidates that trigger installation or runtime errors under the benchmark environment. The resulting set comprises evolution tasks with verifiable behavioral deltas and stable execution characteristics.

\begin{figure*}[t]
        \centering
        \includegraphics[width=0.9\textwidth]{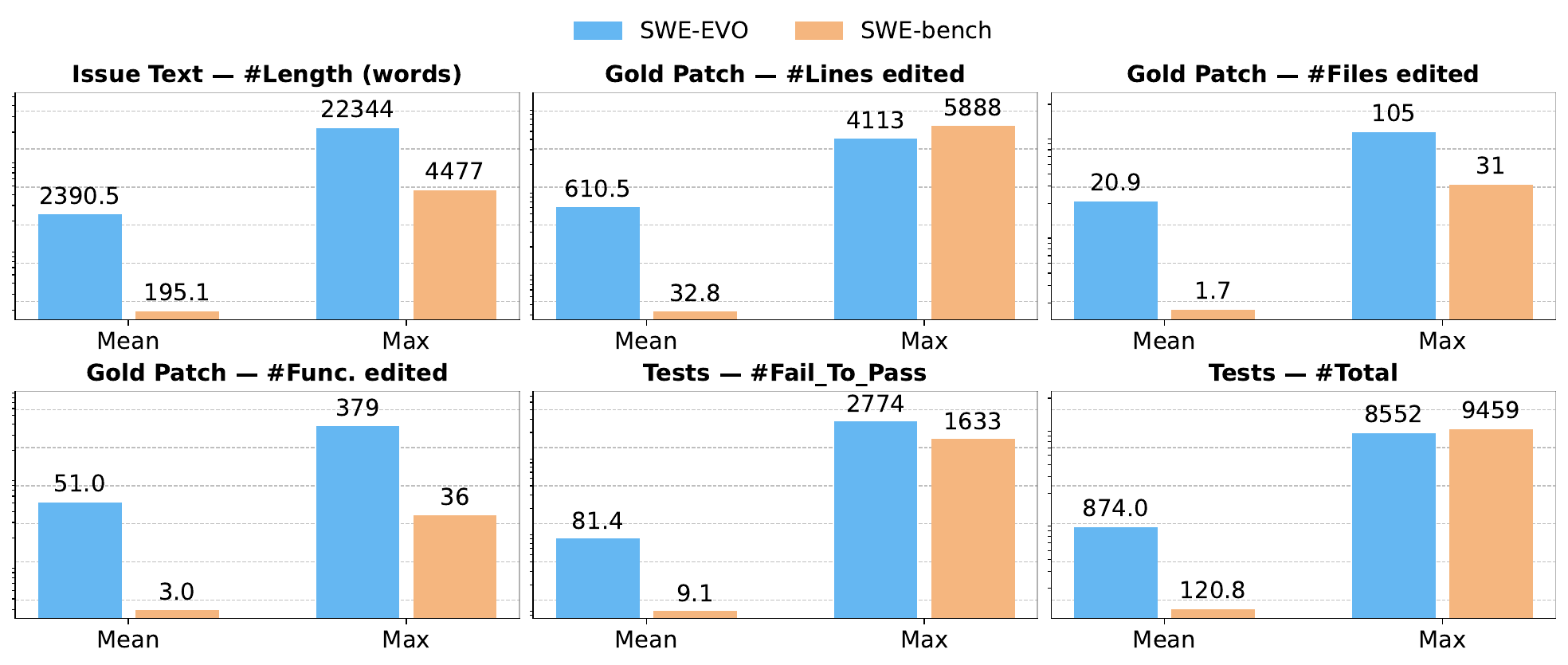}
        \caption{\tool is markedly more demanding than SWE-Bench, with longer specifications, broader patch scope, and heavier test suites in both mean and max.}
        \label{fig:compare_swe_bench}
\end{figure*}

\textcolor{blue}{After the automatic filtering steps yielded a few hundred candidate release transitions, we used human effort to manually verify task quality and scale the benchmark down to 48 high-confidence instances.} This quality-first pass checks alignment among the release note, linked PR/issue context, gold patch, and executable tests. Figure~\ref{fig:repo_dist} reports their distribution across repositories, while Table~\ref{tab:swe_stats} summarizes key characteristics. Compared with SWE-Bench, \tool has longer specifications, broader patches, and heavier test suites (Figure~\ref{fig:compare_swe_bench}), making each human-curated release transition substantially larger in engineering scope than a single-issue repair task. We therefore treat benchmark scale along two axes: number of instances and scope per instance. \tool is smaller on the former, but much larger on the latter, so we avoid over-interpreting close leaderboard gaps while using these release-sized tasks as a stress test of software-evolution capability.

Figure~\ref{fig:compare_swe_bench} makes this scale difference explicit. Across specification length, edited lines, edited files, edited functions, \texttt{FAIL\_TO\_PASS} tests, and total tests, \tool is consistently larger than SWE-Bench in both average and worst-case size. These differences explain why \tool stresses release-level software evolution: agents must interpret broader requirements, coordinate changes across more code, and satisfy substantially heavier verification suites.



\subsection{Task Formulation}
\label{subsec:task_formular}
Having described how \tool instances are constructed, we now formalize the task definition and evaluation metrics.
\paragraph{Model input.} A model is provided with (i) a release-note-centered specification of the intended change (bug fix or feature refinement), consisting of the release note itself and, in our default setting, the text of any PRs or issues explicitly linked from that release note (Appendix~\ref{subsec:PR_Issue_Context}), and (ii) a complete codebase at the \emph{pre-release} commit. The model must produce edits to the codebase that implement the described change. In practice, we represent a solution as a patch file that specifies which lines to modify across files. For clarity, we refer to the patch generated by the model as the \texttt{model patch}, while the ground-truth patch extracted from the start-version to end-version change inside the instance is referred to as the \texttt{gold patch}.

\paragraph{Evaluation Metrics.} Consistent with the evaluation setup of SWE-bench~\citep{jimenez2024swebench}, we use the \textbf{Resolved Rate (\%)} as the primary metric, representing the proportion of task instances successfully solved by the agent. In addition, we report the \textbf{Patch Apply Rate (\%)}, which measures the percentage of generated patches that are syntactically valid and can be applied to the codebase without errors. Beyond the hard-score \textbf{Resolved Rate (\%)}, we introduce a soft-score, \textbf{Fix Rate (\%)}, to provide a more fine-grained assessment of \tool\ while remaining consistent with Resolved Rate.

\textit{Resolved Rate.} Following SWE-Bench, this metric focuses on two categories in the test outcomes:

\begin{itemize}[leftmargin=*]
    \item \texttt{FAIL\_TO\_PASS} tests that were initially failing (\texttt{FAILED} or \texttt{ERROR}) but pass (\texttt{PASSED} after applying the gold patch. 
    \item \texttt{PASS\_TO\_PASS} tests pass both before and after the gold patch and serve as regression checks to ensure unrelated functionality is preserved.
\end{itemize}

The \textbf{Resolved Rate} of an instance is equal $1$ if \emph{all} tests in both \texttt{FAIL\_TO\_PASS} and \texttt{PASS\_TO\_PASS} are \texttt{PASSED}; otherwise, it is $0$. Hence, this metric imposes a strict binary criterion: an instance is counted as resolved only if every relevant test succeeds.

\textit{Fix Rate: A Soft Metric.} While Resolved Rate provides a clear pass/fail signal, it may hide meaningful partial progress, especially when instances contain thousands of tests (Table~\ref{tab:swe_stats}, Figure~\ref{fig:compare_swe_bench}). In such cases, models may fix many failing tests without fully resolving the instance. To capture this progress, we introduce \textbf{Fix Rate}, a soft metric that measures the fraction of \texttt{FAIL\_TO\_PASS} tests successfully fixed.

Fix Rate also enforces a regression constraint: if any \texttt{PASS\_TO\_PASS} test fails after applying the patch, the instance receives a score of $0$. This rewards partial fixes while penalizing regressions. We use this constraint because a valid evolution should implement the requested change without breaking existing behavior. Let $F_i$ denote the \texttt{FAIL\_TO\_PASS} tests and $P_i$ the \texttt{PASS\_TO\_PASS} tests for instance $i$. The \textbf{Fix Rate} is defined in Equation~\ref{eq:fix_rate} as:

\begingroup
\setlength{\abovedisplayskip}{4pt}
\setlength{\belowdisplayskip}{4pt}
\setlength{\abovedisplayshortskip}{4pt}
\setlength{\belowdisplayshortskip}{4pt}
{\small
\begin{equation}
\label{eq:fix_rate}
\mathrm{Fix\ Rate}(i) =
\begin{cases}
\frac{\#\{t \in F_i \mid t \text{ passes}\}}{|F_i|},
& \text{if all } t \in P_i \text{ pass;}\\
0, & \text{otherwise.}
\end{cases}
\end{equation}
}
\endgroup

The overall \textbf{Fix Rate (\%)} across all $N$ instances is $100 \times \frac{1}{N}\sum_{i=1}^{N}\mathrm{Fix\ Rate}(i)$.

Fix Rate lies in $[0,1]$ and is consistent with Resolved Rate: an instance is resolved when its Fix Rate equals $1$. Aggregated across instances, this metric captures partial progress that binary Resolved Rate cannot reflect, while remaining aligned with it (Table~\ref{tab:fixrate_results}). We keep strict Fix Rate as the main soft metric because valid software evolution should make progress without regressing existing behavior. For diagnostics, the same test logs can also report relaxed \texttt{FAIL\_TO\_PASS} progress and \texttt{PASS\_TO\_PASS} preservation separately, distinguishing incomplete implementations from regressions (Appendix~\ref{sec:pass_fail_rate_appendix}). Fix Rate is therefore an execution-based progress signal, not a complete measure of code quality, maintainability, or patch minimality.

\subsection{Features of \tool}

\tool differs from SWE-Bench along several key dimensions. First, it presents substantially richer supervision: longer specifications, broader patches (more lines, files, and functions edited), and heavier test suites (Figure~\ref{fig:compare_swe_bench}). Second, unlike SWE-Bench where each task maps to a single pull request, \tool instances may aggregate multiple PRs, yielding a wide range of difficulty levels (see Appendix~\ref{sec:pr_analysis_appendix}). Finally, evaluation is robust: at least one \texttt{FAIL\_TO\_PASS} test validates each gold patch (81\% have two or more), with an additional mean of 793 \texttt{PASS\_TO\_PASS} regression tests per instance.

%% file: sections/experiments.tex
\section{Experiments}
\label{sec:experiments}

\subsection{Experiment Setup}
We evaluate \tool with two coding-agent scaffolds and a diverse model suite covering frontier proprietary systems and open-weight models.

\paragraph{Agents and Model Selection.} We evaluate \tool using two established coding agent frameworks: \textbf{OpenHands}~\citep{wang2024openhands} (with CodeActAgent, max 100 iterations) and \textbf{SWE-agent}~\citep{yang2024swe} (max 100 LLM calls). We test 18 state-of-the-art LLMs spanning five providers: OpenAI (10 models including GPT-5 series, o3, GPT-4.1, GPT-4o, and GPT-oss-120b), DeepSeek (3 models), Zhipu AI (3 models), Qwen (1 model), and Moonshot AI (2 models). The full list of models with references is provided in Table~\ref{tab:model_list} in the Appendix.



For all reasoning models, we use a medium reasoning effort setting, which balances inference cost and accuracy.

\subsection{Performance on \tool}


Each \tool task is specified by a release-note entry, optionally augmented with linked PR/issue text. To prevent models from retrieving future code or online PR content during evaluation, we block internet access and provide the linked PR/issue text directly in our default \emph{release-note + PR/issue context} setting (Appendix~\ref{subsec:PR_Issue_Context}). Table~\ref{tab:main_results_with_context} reports results for OpenHands and SWE-agent, alongside each model's SWE-bench Verified score when available.

\begin{table*}[ht]
\centering
\small
\setlength{\tabcolsep}{2.5pt}
\renewcommand{\arraystretch}{1.15}
\begin{tabular}{llcc|cc|c}
\toprule
 &  & \multicolumn{2}{c}{\textbf{\tool}} & \multicolumn{2}{c}{\textbf{\tool}} & \textbf{SWE-Bench Verified} \\
\textbf{} & \textbf{Model} & \multicolumn{2}{c}{\textbf{(OpenHands)}} & \multicolumn{2}{c}{\textbf{(SWE-agent)}} & \textbf{(bash only)} \\
\cmidrule(lr){3-4}\cmidrule(lr){5-6}\cmidrule(lr){7-7}
 &  & \% Resolved & \% Apply & \% Resolved & \% Apply & \% Resolved \textsuperscript{\textdagger{}}\\
\midrule
\multirow{10}{*}{OpenAI}
& gpt-5.4 & \textbf{25.00} & 97.92 & \textbf{25.00} & 97.92 & --- \\
& gpt-5.2 & 18.75 & 100 & 22.92 & 97.92 & 72.80 \\
& gpt-5-08-07 & 18.75 & 100 & 20.83 & 100 & 65.00 \\
& gpt-5-mini-08-07 & 10.42 & 97.92 & 10.42 & 100 & 59.80 \\
& gpt-5-nano-08-07 & 4.17 & 85.42 & 4.17 & 100 & 34.80 \\
& o3-2025-04-16 & 4.17 & 93.75 & 6.25 & 100 & 58.40 \\
& gpt-4.1-2025-04-14 & 2.08 & 87.50 & 10.42 & 97.92 & 39.58 \\
& gpt-4o-2024-11-20 & 6.25 & 97.92 & 6.25 & 100 & 21.62 \\
& gpt-oss-120b & 2.08 & 100 & 6.25 & 100 & 26.00 \\
\midrule
\multirow{3}{*}{DeepSeek}
& deepseek-v3p2 & 20.83 & 95.83 & 23.40 & 95.83 & 70.00 \\
& deepseek-v3p1 & 16.67 & 97.92 & 10.42 & 100 & --- \\
& deepseek-r1-0528 & 10.42 & 100 & 8.33 & 100 & 57.60 \\
\midrule
\multirow{1}{*}{Zhipu AI}
& glm-5 & 8.33 & 97.92 & 37.50 & 100 & 72.80 \\
& glm-4p7 & 4.17 & 100 & 39.58 & 97.92 & --- \\
& glm-4p5 & 16.67 & 97.92 & 16.67 & 100 & 54.20 \\
\midrule
Qwen
& qwen3-coder-480b-a35b & 14.58 & 97.92 & 14.58 & 97.92 & 55.40 \\
\midrule
\multirow{2}{*}{Moonshot AI}
& kimi-k2p5 & 22.92 & 97.92 & 25.00 & 97.92 & 70.80 \\
& kimi-k2-instruct & 16.67 & 100 & 18.75 & 100 & 43.80 \\
\bottomrule
\end{tabular}
\caption[]
{Results on \tool with \emph{release note + PR/issue context}: Resolved and Apply rates for OpenHands and SWE-agent. \textsuperscript{\textdagger{}}Results reported on swebench.com~\citep{jimenez2024swebench}. "---" indicates that the result is not reported on the SWE-Bench website.}
\label{tab:main_results_with_context}
\end{table*}

\begin{figure}[ht]
    \centering
    \begin{subfigure}[ht]{0.49\textwidth}
        \centering
        \includegraphics[width=\linewidth]{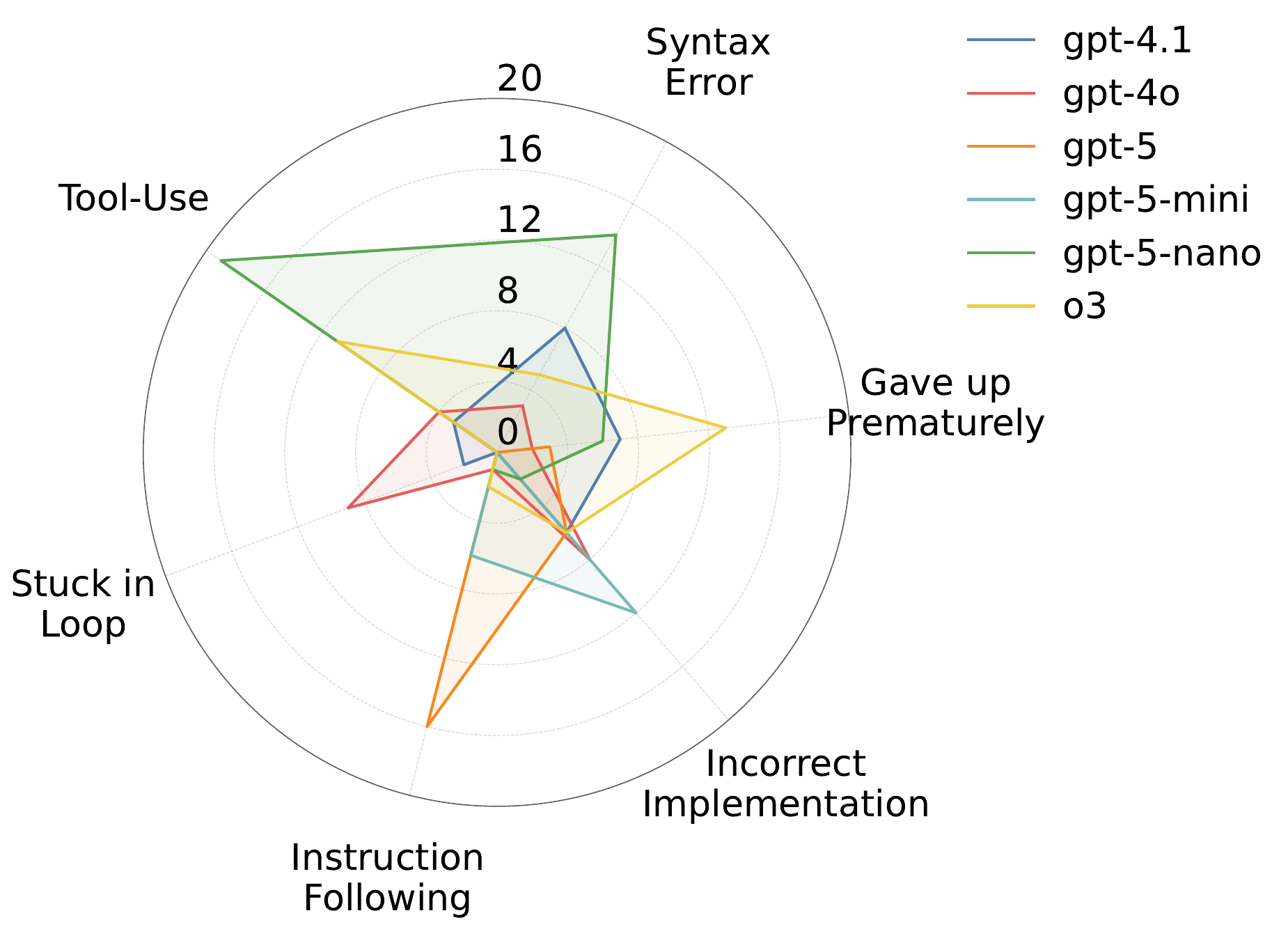}
        \caption{GPT-series models.}
        \label{fig:failure_mode_gpt}
    \end{subfigure}\hfill
    \begin{subfigure}[h]{0.49\textwidth}
        \centering
        \includegraphics[width=\linewidth]{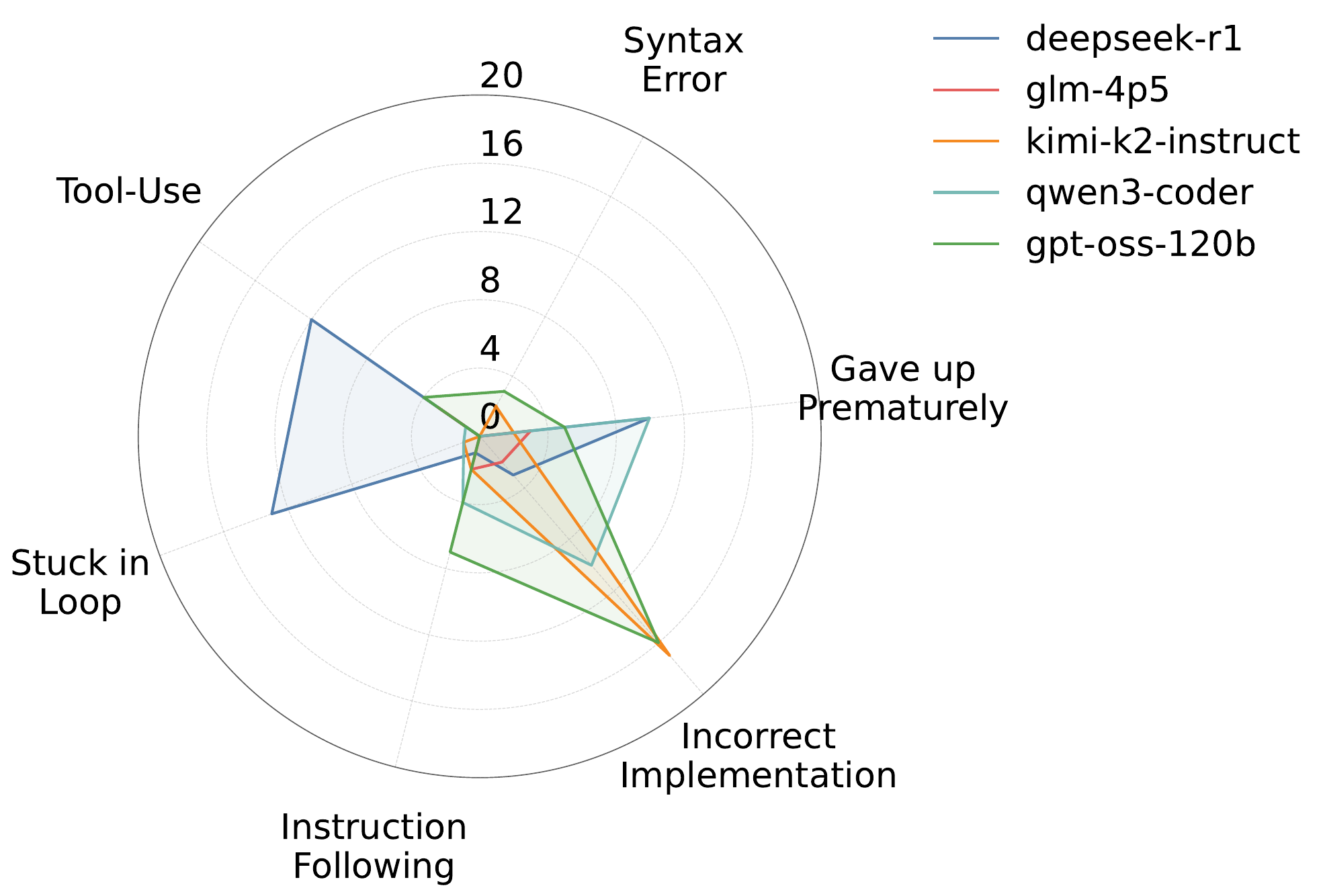}
        \caption{Open-source models.}
        \label{fig:failure_mode_open}
    \end{subfigure}
    \caption{Failure mode distribution for SWE-agent with several model trajectories of unresolved instances. Each instance is automatically labeled using \texttt{gpt-5-mini}~\cite{openai_gpt5} with categories from Table~\ref{tab:failure_mode_descriptions}.}
    \label{fig:failure_mode}
\end{figure}

The scale differences in Figure~\ref{fig:compare_swe_bench} translate into a large performance gap. Even the strongest model \texttt{gpt-5.4} resolves only around \textbf{25\%} of \tool instances, and matched-model comparisons show the same pattern: \texttt{gpt-5.2} drops from 72.80\% on SWE-bench Verified to 22.92\% on \tool. Performance still follows intuitive scaling trends across model families. One exception is \texttt{glm-5}/\texttt{glm-4p7}, which perform much better with SWE-agent than OpenHands, suggesting scaffold sensitivity in addition to benchmark difficulty. Taken together, these results indicate that \tool is a more demanding and more realistic benchmark for end-to-end software-evolution capabilities.

We also evaluate a more challenging \emph{release-note only} setting, where agents receive no PR/issue context. As shown in Table~\ref{tab:main_results_context_compare} (in Appendix~\ref{sec:context_comparison_appendix}), performance drops modestly but trends remain consistent, confirming that \tool poses substantial challenges regardless of specification detail.

\paragraph{Uncertainty from Benchmark Scale.}
Because \tool contains 48 curated instances, one resolved instance changes Resolved Rate by 2.08 percentage points. We therefore report uncertainty and avoid over-interpreting close leaderboard gaps. At the same time, the larger behavioral deltas in Figure~\ref{fig:compare_swe_bench} make each instance closer to a release-sized evolution task than to a single-issue repair task. Representative 95\% Wilson confidence intervals over task instances are [14.9, 38.8] for 25.00\%, [10.2, 31.9] for 18.75\%, [4.5, 22.2] for 10.42\%, and [0.4, 10.9] for 2.08\%, so we focus on large performance gaps rather than small ranking differences.

\paragraph{Fine-Grained Analysis with Fix Rate.}
Beyond the binary \emph{Resolved Rate}, we examine our soft metric \emph{Fix Rate} (Section~\ref{subsec:task_formular}), which captures partial progress on large test suites. As shown in Table~\ref{tab:fixrate_results} (Appendix~\ref{sec:fixrate_appendix}), models indistinguishable under Resolved Rate can differ meaningfully: e.g., under OpenHands, both \texttt{gpt-4.1} and \texttt{gpt-oss-120b} resolve only $2.08\%$ of \tool, but their Fix Rates are $4.65\%$ vs.\ $2.08\%$, revealing that \texttt{gpt-4.1} repairs more failing tests per instance.

\subsection{Analysis of Agent Behaviour}

Beyond aggregate metrics, we investigate \emph{why} agents fail on \tool by analyzing their execution trajectories. This qualitative analysis reveals distinct failure patterns across different model families.

\subsubsection{Trajectory Failure Modes Analysis}


We perform an LLM-as-a-judge analysis of unresolved SWE-agent trajectories, following prior work on SWE-agent~\citep{yang2024swe}. The judge assigns one primary label from the taxonomy in Table~\ref{tab:failure_mode_descriptions}; these labels are qualitative diagnostics rather than execution-based scores, and human agreement validation remains future work. The taxonomy separates low-level execution failures from semantic failures: \emph{Syntax Error} and \emph{Tool-Use} capture invalid patches or failed interaction with the agent tools; \emph{Incorrect Implementation} and \emph{Instruction Following} capture wrong logic or misread requirements; \emph{Stuck in Loop} and \emph{Gave Up Prematurely} capture unproductive or prematurely terminated trajectories; and \emph{Other} covers rare ambiguous cases.

\paragraph{Results.} Figure~\ref{fig:failure_mode} suggests distinct failure patterns across model families. For the GPT series, \texttt{gpt-5} rarely fails due to syntax or tool issues; instead, over 60\% of failures come from \emph{Instruction Following}, suggesting difficulty interpreting complex release notes. Smaller variants (\texttt{gpt-5-mini}, \texttt{gpt-5-nano}) show increasing \emph{Incorrect Implementation}, \emph{Tool-Use}, and \emph{Syntax Error} failures, while older models (\texttt{o3}, \texttt{gpt-4.1}, \texttt{gpt-4o}) exhibit more looping and early-termination issues. In contrast, open and hybrid models follow different patterns: \texttt{kimi-k2-instruct}, \texttt{qwen3-coder}, and \texttt{gpt-oss-120b} fail mainly due to \emph{Incorrect Implementation}, indicating weaker semantic reasoning but stable tool use, whereas \texttt{deepseek-r1} often gets stuck in loops or fails in execution, and \texttt{glm-4p5} shows more evenly distributed errors across categories.

\subsubsection{Difficulty Analysis}
\label{subsubsec:difficulty_analysis}

\begin{figure*}[t]
    \centering
    \begin{subfigure}[t]{0.36\textwidth}
        \centering
        \includegraphics[width=\textwidth]{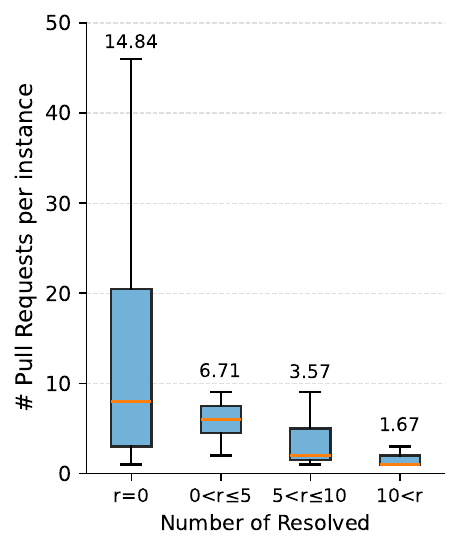}
        \caption{Average number of pull requests per difficulty group.}
        \label{fig:prs_difficult}
    \end{subfigure}
    \hspace{0.03\textwidth}
    \begin{subfigure}[t]{0.58\textwidth}
        \centering
        \includegraphics[width=\textwidth]{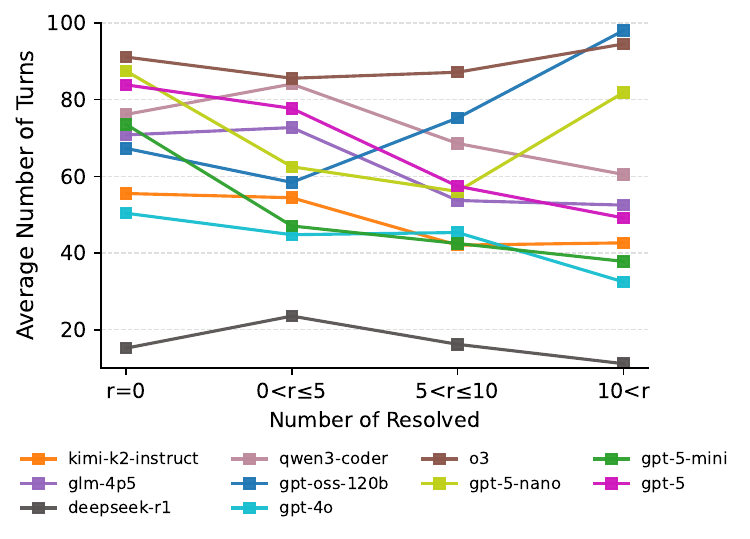}
        \caption{Average number of turns by model across difficulty groups.}
        \label{fig:linecode_difficult}
    \end{subfigure}
    \caption{Difficulty analysis of \tool instances: (a) instances associated with more pull requests are harder; (b) stronger models scale turn usage with difficulty, while others show limited adaptivity.}
    \label{fig:difficulty_combined}
\end{figure*}

A key property of \tool is that instance difficulty varies widely across multiple axes: PR count, specification length, files and lines edited, functions edited, and test-suite size (Table~\ref{tab:swe_stats}). Figure~\ref{fig:difficulty_combined} groups instances by empirical difficulty, using the number of model-scaffold combinations that resolve each instance: $r=0$, $0<r\le5$, $5<r\le10$, and $r>10$, covering roughly 64\%, 15\%, 15\%, and 6\% of instances. Panel (a) shows that harder groups have substantially more linked PRs: unresolved-by-all instances average 14.84 PRs, compared with 1.67 for the easiest group. This supports PR count as a useful proxy for release-level complexity. Panel (b) shows that stronger models tend to spend more turns on harder instances and terminate earlier on easier ones, while weaker models show less adaptive turn allocation.



%% file: sections/conclusion.tex
\section{Conclusion}
\label{sec:conclusion}

We introduce \tool, a benchmark for evaluating coding agents on realistic software evolution rather than isolated bug fixing. \tool requires interpreting release-note-centered specifications, performing multi-file changes, and preserving functionality under comprehensive tests, making it substantially more challenging than SWE-Bench. Experiments with OpenHands and SWE-agent across 18 models reveal a large capability gap: the best model (\texttt{gpt-5.4}) solves only 25\% of tasks, while a matched model such as \texttt{gpt-5.2} drops from 72.80\% on SWE-Bench Verified to 22.92\% on \tool. An exploratory failure analysis suggests that stronger models struggle more with instruction following, while weaker ones exhibit tool-use and syntax errors. We expect \tool to complement existing benchmarks with a focus on long-horizon software evolution.


\section{Limitations}
Our work has some limitations. First, \tool currently covers Python library projects only; this choice keeps the benchmark reproducible and compatible with SWE-Bench/SWE-Gym environments, but multilingual and downstream-application evolution remain future work. Second, our task specifications are release-note-centered, optionally augmented with linked PR/issue text, and therefore do not cover all evolution drivers, such as security advisories, dependency updates, or design-document-driven refactors. Third, the benchmark has 48 curated instances and an imbalanced repository distribution, especially toward \texttt{dvc}. We deliberately prioritize large, execution-validated release transitions over a larger number of shallow tasks, but the current size still limits statistical power for fine-grained comparisons; we therefore provide repository-composition details in Appendix~\ref{sec:per_repo_appendix}. Finally, Fix Rate is execution-verifiable but still weights tests equally and does not measure code quality, maintainability, or patch minimality. We plan to release task metadata, construction scripts, Docker execution settings, generated patches, and aggregate result files to support reproducibility.

%% file: sections/appendix.tex

\section{Evaluated Models}
\label{sec:model_list_appendix}
We evaluate \tool on a diverse set of recent LLMs from multiple providers, including both closed- and open-weight models. Table~\ref{tab:model_list} lists all evaluated models.
\begin{table*}[ht]
\centering
\small
\renewcommand{\arraystretch}{1.15}
\begin{tabular}{llll}
\toprule
\textbf{Provider} & \textbf{Model} & \textbf{Type} & \textbf{Reference} \\
\midrule
\multirow{10}{*}{OpenAI}
& gpt-5.4 & Closed-weight & \citep{openai_gpt54_2026} \\
& gpt-5.2 & Closed-weight & \citep{openai_gpt52_2025} \\
& gpt-5-08-07 & Closed-weight & \citep{openai_gpt5_2025} \\
& gpt-5-mini-08-07 & Closed-weight & \citep{openai_gpt5_2025} \\
& gpt-5-nano-08-07 & Closed-weight & \citep{openai_gpt5_2025} \\
& o3-2025-04-16 & Closed-weight & \citep{openai_o3_2025} \\
& gpt-4.1-2025-04-14 & Closed-weight & \citep{openai_gpt41_2025} \\
& gpt-4o-2024-11-20 & Closed-weight & \citep{openai_gpt4o_2024} \\
& gpt-oss-120b & Open-weight & \citep{openai_gptoss_2025} \\
\midrule
\multirow{3}{*}{DeepSeek}
& deepseek-v3p2 & Open-weight & \citep{deepseek_v32_2025} \\
& deepseek-v3p1 & Open-weight & \citep{deepseek_v31_2025} \\
& deepseek-r1-0528 & Open-weight & \citep{deepseek_r1_0528_2025} \\
\midrule
\multirow{3}{*}{Zhipu AI}
& glm-5 & Open-weight & \citep{zai_glm5_2026} \\
& glm-4p7 & Open-weight & \citep{zai_glm47_2025} \\
& glm-4p5 & Open-weight & \citep{glm45team_2025} \\
\midrule
Qwen
& qwen3-coder-480b-a35b & Open-weight & \citep{qwen3coder_2025,qwen3_2025} \\
\midrule
\multirow{2}{*}{Moonshot AI}
& kimi-k2p5 & Open-weight & \citep{kimi_k25_2026} \\
& kimi-k2-instruct & Open-weight & \citep{kimi_k2_2025} \\
\bottomrule
\end{tabular}
\caption{Full list of LLMs evaluated on \tool, grouped by provider.}
\label{tab:model_list}
\end{table*}

\section{Software Evolution Conceptual Model}
\label{sec:evolution_model_appendix}

\begin{figure*}[ht]
        \centering
        \includegraphics[width=0.7\textwidth]{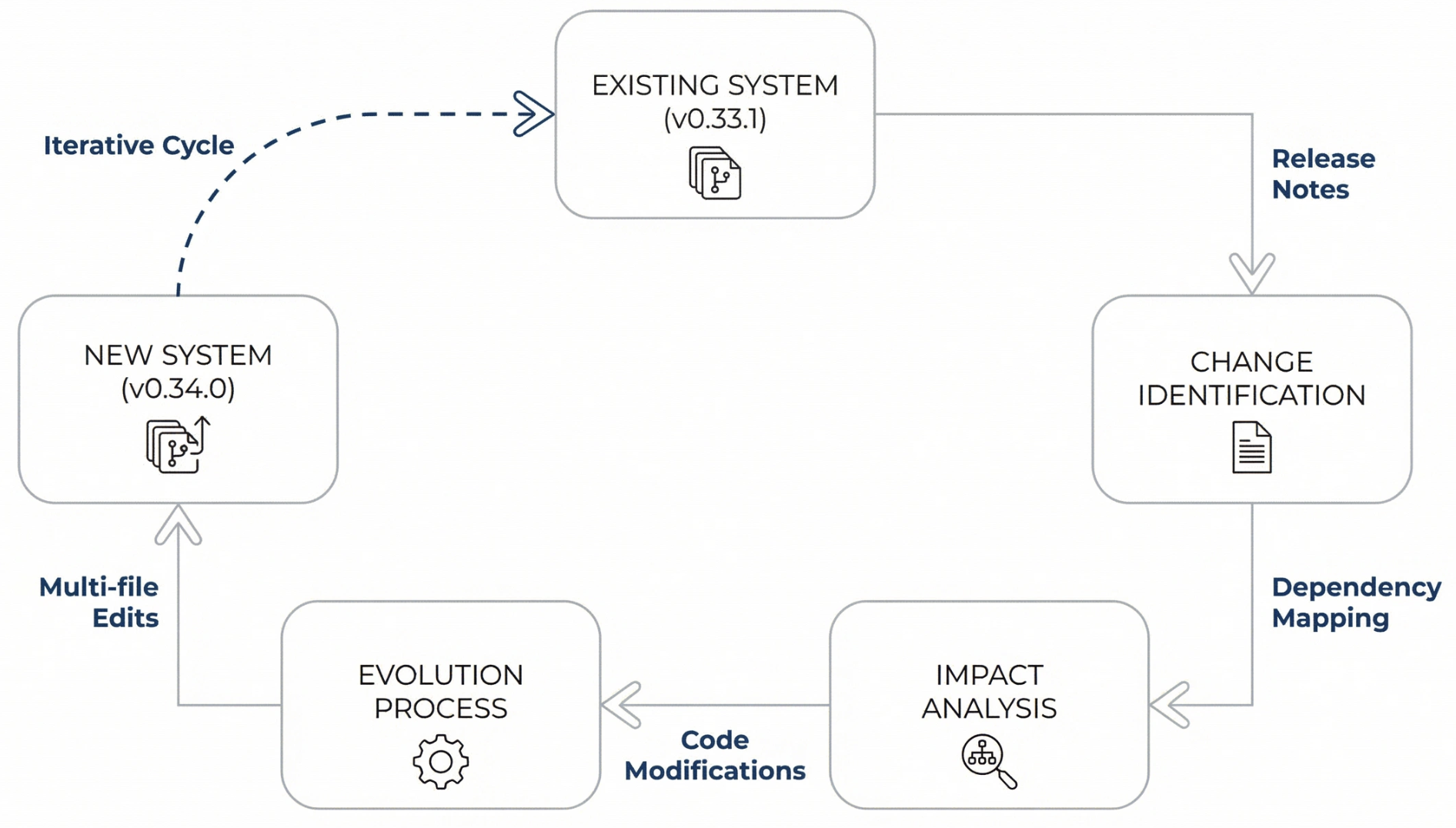}
        \caption{Conceptual model of software evolution, depicting the iterative cycle from an existing system to a new system through change identification, impact analysis, and the evolution process.}
        \label{fig:software-evolution}
\end{figure*}

Software evolution is an iterative cycle where, starting from an existing system, engineers identify required changes, analyze their impact, and carry out an evolution process that yields a new system aligned with updated requirements. This cycle repeats as the new system becomes the starting point for subsequent iterations. \tool captures this process by tasking agents with evolving codebases between consecutive release versions.

\section{Self-Evolving Agents and Tool Creation}
\label{sec:self_evolving_appendix}

Due to the huge design space for software agents, building an optimal agent scaffold can be extremely challenging and costly. As a result, several self-improving and self-evolving software agents have emerged recently. The Self-Improving Coding Agent (SICA)~\citep{robeyns2025self} introduced mechanisms for agents to learn from their own experiences. Darwin-G{\"o}del Machine (DGM)~\citep{zhang2025darwin} proposed open-ended evolution of self-improving agents through iterative refinement. Huxley-G{\"o}del Machine (HGM)~\citep{wang2025huxley} further advanced this paradigm by approximating optimal self-improving machines for human-level coding. However, such self-improving agents typically require costly offline training on known benchmarks and may not generalize well across different LLMs, benchmarks, and issue types.

Beyond the software engineering domain, prior work has explored using LLMs to create tools for general reasoning or embodied tasks. Large Language Models as Tool Makers (LATM)~\citep{llmastool} demonstrated that LLMs can autonomously create reusable tools to solve complex tasks more efficiently. Voyager~\citep{wang2024voyager} introduced an open-ended embodied agent that continuously acquires new skills through code generation in Minecraft. CREATOR~\citep{qian2024creatortoolcreationdisentangling} proposed disentangling abstract and concrete reasoning through tool creation. TroVE~\citep{wang2024troveinducingverifiableefficient} focused on inducing verifiable and efficient toolboxes for programmatic tasks. Alita~\citep{qiu2025alitageneralistagentenabling} presented a generalist agent enabling scalable agentic reasoning with minimal predefinition and maximal self-evolution. While these works demonstrate the potential of self-evolving agents and tool creation, they do not specifically target real-world software engineering problems that require long-horizon evolution across multiple commits and versions.

\section{SWE-EVO vs.\ SWE-Bench Comparison}
\label{sec:swebench_comparison_appendix}
The main paper reports the quantitative comparison in Figure~\ref{fig:compare_swe_bench}. The same statistics are derived from instance-level metadata for \tool and SWE-Bench, covering specification length, gold-patch scope, and test-suite size.

\section{Context Comparison Results}
We further compare performance under \emph{release-note only} and \emph{release-note + PR/issue context} settings. As shown in Table~\ref{tab:main_results_context_compare}, adding PR/issue context generally improves resolved rates across most models, while preserving the overall ranking and relative trends. This suggests that additional context provides useful signals.
\label{sec:context_comparison_appendix}
\begin{table*}[ht]
\centering
\small
\setlength{\tabcolsep}{4pt}
\renewcommand{\arraystretch}{1.15}
\begin{tabular}{lcc|cc}
\toprule
& \multicolumn{2}{c|}{\shortstack{\textbf{release-note only}}}
& \multicolumn{2}{c}{\shortstack{\textbf{release-note} \\ \textbf{+ PR/issue context}}} \\
\cmidrule(lr){2-3}\cmidrule(lr){4-5}
\textbf{Model} & \text{OpenHands} & \text{SWE-agent} & \text{OpenHands} & \text{SWE-agent} \\
\midrule
gpt-5-08-07             & \textbf{14.58} & \textbf{16.67} & \textbf{18.75} & \textbf{20.83} \\
gpt-5-mini-08-07        & 8.33           & 10.42          & 10.42 & 10.42 \\
o3-2025-04-16           & 4.17           & 6.25           & 4.17 & 6.25 \\
gpt-4.1-2025-04-14      & 2.08           & 8.33           & 2.08 & 10.42 \\
gpt-4o-2024-11-20       & 8.33           & 4.17           & 6.25 & 6.25 \\
gpt-oss-120b            & 0.00           & 0.00           & 2.08 & 6.25 \\
deepseek-r1-0528        & 10.42          & 6.25           & 10.42 & 8.33 \\
glm-4p5                 & 6.25           & 12.50          & 16.67 & 16.67 \\
qwen3-coder-480b-a35b   & \textbf{14.58} & 12.50          & 14.58 & 14.58 \\
kimi-k2-instruct        & 8.33           & 12.50          & 16.67 & 18.75 \\
\bottomrule
\end{tabular}
\caption{Results on \tool under two settings: \textbf{release-note only} and \textbf{release-note + PR/issue context}. Reported metric is \textbf{Resolved Rate (\%)} on \textbf{OpenHands} and \textbf{SWE-agent}.}
\label{tab:main_results_context_compare}
\end{table*}

\section{Pull Request Distribution Analysis}
\label{sec:pr_analysis_appendix}

\begin{figure*}[ht]
    \centering
    \begin{subfigure}[t]{0.32\textwidth}
        \centering
        \includegraphics[width=\linewidth]{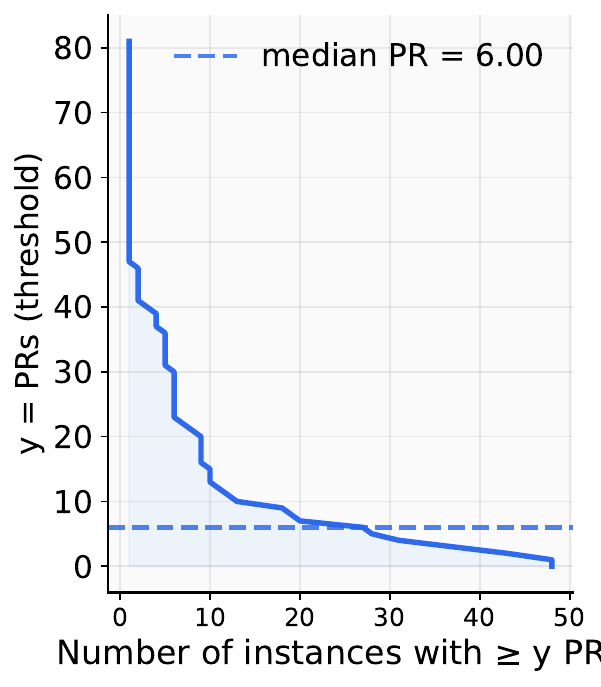}
        \caption{Complementary cumulative distribution of PR counts, showing how many instances contain at least $y$ linked pull requests.}
        \label{fig:ccdf_prs}
    \end{subfigure}\hfill
    \begin{subfigure}[t]{0.65\textwidth}
        \centering
        \includegraphics[width=\linewidth]{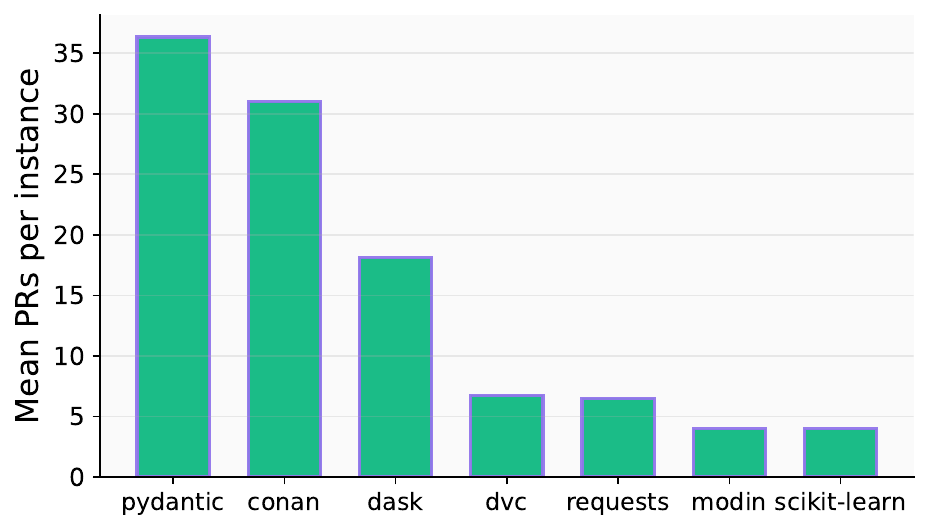}
        \caption{Mean PRs per repository, highlighting that certain codebases require substantially more upstream context to resolve.}
        \label{fig:prs_repo_dist}
    \end{subfigure}
    \caption{Pull Request statistics across \tool{} instances.}
    \label{fig:PRs_dist}
\end{figure*}

Unlike SWE-Bench, where each task instance corresponds to a single pull request (PR), an instance in \tool may be associated with multiple pull requests that collectively implement or refine a release-note change. We hypothesize that instances associated with a larger number of pull requests reflect more complex, multi-step development efforts, as each pull request typically represents a distinct feature enhancement or bug fix. As shown in Figure~\ref{fig:PRs_dist}, the number of pull requests per instance varies widely, indicating that \tool spans a broad range of difficulty levels.

\section{Difficulty Analysis}
\label{sec:difficulty_appendix}

The main paper reports the difficulty analysis in Figure~\ref{fig:difficulty_combined} and Section~\ref{subsubsec:difficulty_analysis}. We keep this appendix section as a pointer so the appendix PR-distribution analysis remains connected to the promoted main-text result without duplicating the same figure.

\section{Repository Composition and Bias Checks}
\label{sec:per_repo_appendix}

Table~\ref{tab:repo_performance} reports per-repository Resolved Rate and Fix Rate, averaged over these evaluated models: \texttt{gpt-5-08-07}, \texttt{gpt-5-mini-08-07}, \texttt{gpt-5-nano-08-07}, \texttt{o3-2025-04-16}, \texttt{gpt-4.1}, \texttt{gpt-4o}, \texttt{gpt-oss-120b}, \texttt{deepseek-r1}, \texttt{glm-4p5}, \texttt{qwen3-coder-480b}, and \texttt{kimi-k2-instruct}. Our strict collection process requires stable release tags, reproducible environments, and executable tests, which improves validity but introduces some quantity imbalance: 26 of 48 instances come from \texttt{dvc}. However, the resulting metrics are not concentrated on a single repository. Performance remains low across repositories, and the highest aggregate values vary by scaffold. We therefore treat repository composition as a dataset caveat, but the metric itself does not appear to favor any single repository.

\begin{table*}[ht]
\centering
\small
\setlength{\tabcolsep}{5pt}
\renewcommand{\arraystretch}{1.1}
\begin{tabular}{lrr|cc|cc}
\toprule
\multirow{2}{*}{\textbf{Repository}} & \multirow{2}{*}{\textbf{\# Inst.}} & \multirow{2}{*}{\textbf{Share (\%)}} &
\multicolumn{2}{c|}{\textbf{OpenHands}} &
\multicolumn{2}{c}{\textbf{SWE-agent}} \\
\cmidrule(lr){4-5} \cmidrule(lr){6-7}
& & & Resolved & Fix & Resolved & Fix \\
\midrule
\texttt{dvc} & 26 & 54.2 & 4.36 & 6.69 & 8.72 & 5.63 \\
\texttt{dask} & 8 & 16.7 & 2.68 & 1.14 & 2.57 & 0.83 \\
\texttt{requests} & 4 & 8.3 & 2.65 & 2.46 & 3.12 & 2.92 \\
\texttt{pydantic} & 3 & 6.2 & 0.00 & 0.00 & 0.01 & 0.00 \\
\texttt{modin} & 3 & 6.2 & 1.23 & 1.14 & 1.04 & 1.04 \\
\texttt{conan} & 2 & 4.2 & 0.38 & 0.38 & 0.31 & 0.21 \\
\texttt{scikit-learn} & 2 & 4.2 & 0.19 & 0.19 & 0.00 & 0.00 \\
\midrule
\textbf{Total} & 48 & 100.0 & -- & -- & -- & -- \\
\bottomrule
\end{tabular}
\caption{Repository composition and aggregate per-repository performance. Resolved and Fix values are percentages averaged over these evaluated models for each scaffold.}
\label{tab:repo_performance}
\end{table*}

\section{Fix Rate Results}
\label{sec:fixrate_appendix}
Beyond the binary \emph{Resolved Rate}, we also examine our soft metric \emph{Fix Rate}, which captures partial progress on large test suites (Section~\ref{subsec:task_formular}). As shown in Table~\ref{tab:fixrate_results}, models that appear similar under Resolved Rate can differ meaningfully once partial improvements are considered. For example, under OpenHands, both \texttt{gpt-4.1} and \texttt{gpt-oss-120b} resolve only 2.08\% of \tool, but their Fix Rates differ (4.65\% vs.\ 2.08\%), indicating that \texttt{gpt-4.1} consistently fixes more failing tests per instance. More broadly, we observe a consistent gap between Fix Rate and Resolved Rate across all models, suggesting that many trajectories make partial but incomplete progress. This finer granularity is especially important on \tool, where each task involves many tests, and binary outcomes would otherwise obscure systematic differences in model capability.

\begin{table*}[h]
\centering
\small
\setlength{\tabcolsep}{2.5pt}
\renewcommand{\arraystretch}{1.1}
\begin{tabular}{llcc|cc|cc}
\toprule
\multirow{2}{*}{\textbf{}} & \multirow{2}{*}{\textbf{Model}} &
\multicolumn{2}{c|}{\textbf{OpenHands}} &
\multicolumn{2}{c|}{\textbf{SWE-agent}} &
\multicolumn{2}{c}{\textbf{Average}} \\
\cmidrule(lr){3-4} \cmidrule(lr){5-6} \cmidrule(lr){7-8}
 & & Resolved (\%) & Fix (\%) & Resolved (\%) & Fix (\%) & Resolved (\%) & Fix (\%) \\
\midrule
\multirow{10}{*}{OpenAI}
& gpt-5.4              & \textbf{25.00} & \textbf{33.89} & \textbf{25.00}    & \textbf{33.98}    & \textbf{25.00}    & \textbf{33.96}    \\
& gpt-5.2              & 18.75    & 23.43    & 22.92 & 30.21 & 20.84    & 26.82    \\
& gpt-5-08-07          & 18.75 & 27.64 & 20.83 & 31.46 & 19.79 & 29.55 \\
& gpt-5-mini-08-07     & 10.42 & 17.48 & 10.42 & 17.48 & 10.42 & 17.48 \\
& gpt-5-nano-08-07     & 4.17  & 5.99  & 4.17  & 5.26  & 4.17  & 5.63  \\
& o3-2025-04-16        & 4.17  & 6.47  & 4.17  & 13.72 & 4.17  & 10.10 \\
& gpt-4.1-2025-04-14   & 2.08  & 4.65  & 10.42 & 14.79 & 6.25  & 9.72  \\
& gpt-4o-2024-11-20    & 6.25  & 7.77  & 6.25  & 10.15 & 6.25  & 8.96  \\
& gpt-oss-120b         & 2.08  & 2.08  & 6.25  & 7.88  & 4.17  & 4.98  \\
\midrule
\multirow{3}{*}{DeepSeek}
& deepseek-v3p2        & 20.83 & 26.89 & 23.40 & 31.41 & 22.12 & 29.15 \\
& deepseek-v3p1        & 16.67 & 21.83 & 10.42 & 15.51 & 13.55 & 18.67 \\
& deepseek-r1-0528     & 10.42 & 14.31 & 8.33  & 9.89  & 9.38  & 12.10 \\
\midrule
\multirow{1}{*}{Zhipu AI}
& glm-5                & 8.33  & 9.67  & 37.50 & 44.27 & 22.92 & 27.45 \\
& glm-4p7              & 4.17  & 5.19  & 39.58 & 45.87 & 21.88 & 25.53 \\
& glm-4p5              & 16.67 & 23.74 & 16.67 & 26.55 & 16.67 & 25.15 \\
\midrule
Qwen
& qwen3-coder-480b-a35b & 14.58 & 19.56 & 14.58 & 23.74 & 14.58 & 21.65 \\
\midrule
\multirow{2}{*}{Moonshot AI}
& kimi-k2p5            & 22.92 & 27.75 & \textbf{25.00} & 32.91 & 23.96 & 30.33 \\
& kimi-k2-instruct     & 16.67 & 22.42 & 18.75 & 24.03 & 17.71 & 23.23 \\
\bottomrule
\end{tabular}
\caption{Comparison of \textbf{Resolved Rate (\%)} and \textbf{Fix Rate (\%)} across models on \textbf{OpenHands}, \textbf{SWE-agent}, and their \textbf{Average}.}
\label{tab:fixrate_results}
\end{table*}

\section{FAIL\_TO\_PASS and PASS\_TO\_PASS Diagnostics}
\label{sec:pass_fail_rate_appendix}

Table~\ref{tab:pass_fail_diagnostics} reports detailed \texttt{FAIL\_TO\_PASS} and \texttt{PASS\_TO\_PASS} rates for each model in these evaluated model and each scaffold. These diagnostics separate two effects that are combined by strict Fix Rate: whether the model implements the requested behavioral change, and whether it preserves tests that already passed before the patch. Across most models, higher \texttt{PASS\_TO\_PASS} rates tend to coincide with lower \texttt{FAIL\_TO\_PASS} rates, suggesting a trade-off: as a model spends more effort resolving the new task, it may be more likely to forget or break existing codebase behavior. However, valid software evolution should satisfy the requested change without breaking existing behavior. We therefore use these rates as diagnostics, while keeping strict Fix Rate as the main soft metric.

\begin{table*}[ht]
\centering
\small
\setlength{\tabcolsep}{4pt}
\renewcommand{\arraystretch}{1.08}
\begin{tabular}{lcc|cc}
\toprule
\multirow{2}{*}{\textbf{Model}} &
\multicolumn{2}{c|}{\textbf{OpenHands}} &
\multicolumn{2}{c}{\textbf{SWE-agent}} \\
\cmidrule(lr){2-3} \cmidrule(lr){4-5}
& \texttt{FAIL\_TO\_PASS} & \texttt{PASS\_TO\_PASS}
& \texttt{FAIL\_TO\_PASS} & \texttt{PASS\_TO\_PASS} \\
\midrule
\texttt{gpt-5-08-07} & 65.68 & 34.32 & 59.95 & 37.97 \\
\texttt{gpt-5-mini-08-07} & 64.15 & 23.35 & 76.27 & 21.65 \\
\texttt{gpt-5-nano-08-07} & 58.59 & 8.07 & 90.58 & 7.34 \\
\texttt{o3-2025-04-16} & 85.20 & 8.55 & 82.11 & 15.81 \\
\texttt{gpt-4.1} & 82.85 & 4.65 & 71.67 & 24.16 \\
\texttt{gpt-4o} & 87.20 & 10.89 & 81.62 & 16.30 \\
\texttt{gpt-oss-120b} & 95.83 & 4.17 & 86.91 & 11.00 \\
\texttt{deepseek-r1} & 77.32 & 22.68 & 83.23 & 14.69 \\
\texttt{glm-4p5} & 68.64 & 29.28 & 62.57 & 35.35 \\
\texttt{qwen3-coder-480b} & 68.54 & 29.38 & 62.97 & 32.86 \\
\texttt{kimi-k2-instruct} & 71.45 & 28.55 & 67.52 & 30.39 \\
\bottomrule
\end{tabular}
\caption{\texttt{FAIL\_TO\_PASS} and \texttt{PASS\_TO\_PASS} rates (\%) by model and scaffold. The diagnostics separate progress on required behavior from preservation of previously passing tests.}
\label{tab:pass_fail_diagnostics}
\end{table*}

\section{Data and Code Availability}
\label{sec:availability_appendix}

\tool is designed for public release. We will release the task metadata, release-note and PR/issue context construction scripts, Docker-based execution configuration, evaluation harness, generated patches, and aggregate result files needed to reproduce the reported metrics. The release will also include the dataset fields described in Table~\ref{tab:instance_fields} and instructions for running OpenHands and SWE-agent under the settings used in this paper. When required for anonymous review, these artifacts should be distributed through an anonymized repository or supplemental archive.

\section{Failure Mode Descriptions}
\label{subsec:Failure_mode_appendix}
We use a coarse taxonomy that separates execution failures, semantic failures, and unproductive trajectories. For each unresolved SWE-agent trajectory, the judge assigns exactly one primary label; when multiple issues are present, it chooses the most fundamental cause. Table~\ref{tab:failure_mode_descriptions} summarizes the labels used in the main analysis.

\begin{table*}[ht]
\centering
\footnotesize
\setlength{\tabcolsep}{3pt}
\renewcommand{\arraystretch}{1.05}
\begin{tabular}{p{0.22\linewidth}|p{0.72\linewidth}}
\toprule
\textbf{Category} & \textbf{Description} \\
\midrule
Syntax Error &
Patch introduces parse, formatting, import, indentation, JSON/YAML, or similar errors that prevent execution. \\
\midrule
Incorrect Implementation &
Patch touches a plausible area but implements the required behavior incorrectly or incompletely. \\
\midrule
Instruction Following &
Agent misreads, ignores, or deviates from the release note or linked requirement, effectively solving the wrong task. \\
\midrule
Tool-Use &
Progress is blocked by failed or incorrect tool use, such as bad edits, missed tests, or wrong paths. \\
\midrule
Stuck in Loop &
Agent repeatedly reads, edits, or reruns tests without substantive progress. \\
\midrule
Gave Up Prematurely &
Agent stops or declares failure while reasonable next steps remain. \\
\midrule
Other &
Rare or ambiguous failure not captured by the other labels. \\
\bottomrule
\end{tabular}
\caption{Descriptions of failure mode categories.}
\label{tab:failure_mode_descriptions}
\end{table*}

\section{Dataset Fields}
Table~\ref{tab:instance_fields} describes the \tool dataset fields and outlines how they are obtained throughout the curation process.

\begin{table*}[ht]
\centering
\small
\begin{tabular}{llp{0.6\textwidth}}
\toprule
\textbf{Field} & \textbf{Type} & \textbf{Description} \\
\midrule
\texttt{repo} & \texttt{str} &
Git repository identifier for the task instance, e.g., the GitHub \texttt{owner/name}. \\

\texttt{base\_commit} & \texttt{str} &
The commit on which the pull request is based, representing the repository state before the issue is resolved. \\

\texttt{start\_version} & \texttt{str} &
Git tag or version identifier of the starting release for this task instance. \\

\texttt{end\_version} & \texttt{str} &
Git tag or version identifier of the target release after the change has been applied. \\

\texttt{end\_version\_commit} & \texttt{str} &
Git commit hash corresponding to the \texttt{end\_version} tag. \\

\texttt{patch} & \texttt{str} &
Gold patch proposed by the pull request, in \texttt{.diff} format. \\

\texttt{test\_patch} & \texttt{str} &
Modifications to the test suite proposed by the pull request that are typically used to check whether the issue has been resolved. \\

\texttt{problem\_statement} & \texttt{str} &
Issue description text, typically describing the bug or requested feature, used as the task problem statement. \\

\texttt{FAIL\_TO\_PASS} & \texttt{List[str]} &
Test cases that are expected to successfully transition from failing to passing and are used to evaluate the correctness of the patch. \\

\texttt{PASS\_TO\_PASS} & \texttt{List[str]} &
Test cases that are already passing prior to applying the gold patch; a correct patch should not introduce regression failures in these tests. \\

\texttt{*image} & \texttt{str} &
Instance-level Docker image that provides an execution environment. \\

\texttt{*test\_cmds} & \texttt{List[str]} &
Command(s) used to run the test suite, as identified by the verify agent in \textsc{RepoLaunch}, enabling detailed logging of each test item's status (e.g., via \texttt{pytest -rA}). \\

\texttt{*log\_parser} & \texttt{str} &
Type of log parser required for the instance---by default, \texttt{pytest}. \\
\bottomrule
\end{tabular}
\caption{Required fields for a typical issue-solving task instance. Fields marked with * are newly added compared to SWE-Bench.}
\label{tab:instance_fields}
\end{table*}

\section{Problem Statement}
\label{subsec:Problem_Statement_Appendix}
For each \tool instance, the problem statement is defined by the official release notes text that describes the changes between the \emph{start} version and the \emph{end} version of this instance, as published by the project maintainers. This release note is then presented to the agent as the sole natural-language specification of how the codebase should change between the two versions.

\subsection{Pull Request, Issue Context}
\label{subsec:PR_Issue_Context}
Release notes frequently refer to specific pull requests or issues (e.g., ``see \#1234'' or ``thanks to PR \#5678''). To make these references actionable for agent, we augment the problem statement with the context of the linked artifacts. For each referenced pull request or issue, we fetch its body from github. These texts are then concatenated into a compact ``PR / Issue Context'' block that is presented to the agent alongside the release note. We then append these texts directly below the release note in a structured format, without rewriting or summarizing them. In general, the final problem statement presented to the agent has the form:
\emph{release-note} followed by a sequence of sections
\texttt{\textbackslash n\#\#\# PR xxx:\textbackslash n} or \texttt{\textbackslash n\#\#\# Issue yyy:\textbackslash n} with their corresponding content for every referenced pull request and issue.
so that the agent sees the original release note followed by the full text of any referenced pull requests and issues.

\subsection{Example}
To make this concrete, we present two example problem statements (Boxes~\ref{box:ex1} and~\ref{box:ex2}), corresponding to the instances \texttt{iterative\_\_dvc\_0.33.1\_0.34.0} in the \texttt{iterative/dvc} repository and \texttt{dask\_\_dask\_2023.6.1\_2023.7.0} in the \texttt{dask/dask} repository.

\clearpage
\onecolumn
\begin{problemstatementboxA}
    \label{box:ex1}
    1) [`dvc metrics show` now nicely formats multiline metrics files like tsv/htsv, csv/hcsv, json](https://github.com/iterative/dvc/issues/1716); Kudos @mroutis :medal\_sports: \\
    2) [`dvc remote add` no longer silently overwrites existing sections](https://github.com/\\
    iterative/dvc/issues/1760);\\
3) [Use a workaround to bypass SIP protection on osx, when accessing libSystem](https://github.com/iterative/dvc/issues/1515);\\
4) [Don't try to create existing container on azure](https://github.com/iterative/dvc/issues/\\
1811); Kudos @AmitAronovitch :medal\_sports: \\
5) Dvc repository now uses read-only http remote for our images instead of s3;\\
6) Fix bug in `dvc status` where an error is raised if cache is not present locally nor on the remote;
7) [Fix progress bar on `dvc pull`]\\(https://github.com/iterative/dvc/issues/1807); Kudos @pared :medal\_sports: \\
8) [Automatically detect metrics file type by extension](https://github.com/iterative/dvc/issues/\\
1553); Kudos @cand126 :medal\_sports: \\

Welcome new contributor @AmitAronovitch ! :tada:\\

\vspace{0em}

\textbf{\#\#\# Issue 1716:}\\
When displaying TSV metrics output the printout is not readable:

\texttt{../../20181223-TrainSetZero/SanityCheck/Bravo\_on\_TrainSetZero.metrics: value\_mse deviation\_mse data\_set \\
0.421601 0.173461 train \\
0.67528 0.289545 testing \\
0.671502 0.297848 validation}

I think it would be much easier to read if a newline were added, and the remaining text were formatted as though it had been passed via \texttt{column -t}:

\texttt{../../20181223-TrainSetZero/SanityCheck/Bravo\_on\_TrainSetZero.metrics: \\
value\_mse  deviation\_mse  data\_set \\
0.421601   0.173461       train \\
0.67528    0.289545       testing \\
0.671502   0.297848       validation}\\

\vspace{0em}

\textbf{\#\#\# Issue 1807:}\\
\texttt{(3.7.0-dvc) \textrightarrow  dvc git:(dvctags) \texttimes dvc pull \\
Preparing to download data from 's3://dvc-share/dvc/' \\
Preparing to collect status from s3://dvc-share/dvc/ \\
{[\#\#\#\#\#\#\#\#\#\#\#\#\#\#\#\#\#\#\#\#\#\#\#\#\#\#\#\#\#\#] 100\% Collecting information} \\
{[\#\#\#\#\#\#\#\#\#\#\#\#\#\#\#\#\#\#\#\#\#\#\#\#\#\#\#\#\#\#] 100\% Analysing status.} \\
(1/3): {[\#\#\#\#\#\#\#\#\#\#\#\#\#\#\#\#\#\#\#\#\#\#\#\#\#\#\#\#\#\#] 100\% dvc\_up.bmp} \\
(2/3): {[\#\#\#\#\#\#\#\#\#\#\#\#\#\#\#\#\#\#\#\#\#\#\#\#\#\#\#\#\#\#] 100\% dvc.ico} \\
(3/3): {[\#\#\#\#\#\#\#\#\#\#\#\#\#\#\#\#\#\#\#\#\#\#\#\#\#\#\#\#\#\#] 100\% dvc\_left.bmp} \\
(4/3): {[\#\#\#\#\#\#\#\#\#\#\#\#\#\#\#\#\#\#\#\#\#\#\#\#\#\#\#\#\#\#] 100\% Checkout finished!}}

looks like checkout didn't reset progress counter\\

\vspace{0em}

\textbf{\#\#\# Issue 1553:}\\
E.g. if we see that output has .json suffix, we could safely assume that it is \texttt{--type json} without explicitly specifying it.\\

\end{problemstatementboxA}
\clearpage
\begin{problemstatementboxA}[title=Problem Statement: iterative\_\_dvc\_0.33.1\_0.34.0 (continued)]

\textbf{\#\#\# Issue 1760:}\\
The \texttt{dvc remote add} command ignores the existing remote and overwrites it silently.

\texttt{\textrightarrow dvc --version \\
0.32.1+7d7ed4}

To reproduce \\
\texttt{dvc remote add s3 s3://bucket/subdir \\
dvc remote add s3 s3://bucket/subdir2}

Expected behavior \\
The second command \texttt{dvc remote add s3 s3://bucket/subdir2} should fail with the Remote with name "s3" already exists message.

Current behavior \\
Remote URL is silently overwriten:

\texttt{> cat .dvc/config \\
{[ 'remote "s3"' ]} \\
url = s3://bucket/subdir2}\\

\vspace{0em}

\textbf{\#\#\# Issue 1811:} \\
Azure SAS connection strings can be used to allow read-only access to specific container, which is useful in the context of CI pipelines and automation.

However, when using such limited-credentials connection string, \texttt{dvc pull} abends with the error below.

Some digging reveals that this is because \texttt{remote/azure.py} always tries to create the bucket (which already exists). If we check for existence before trying to create - this should work...

(will send PR)

\texttt{{[\#\#\#\#\#\#\#\#\#                     ] 30\% Collecting information}} \\
\texttt{Client-Request-ID=19129d6a-5393-11e9-80ab-5800e34ea0d9}\\
\texttt{Retry policy did not allow for a retry:}\\
\texttt{Server-Timestamp=Sun, 31 Mar 2019 08:58:06 GMT,}\\
\texttt{Server-Request-ID=c72fbeda-501e-0089-3c9f-e78298000000,}\\
\texttt{HTTP status code=403, Exception=This request is not}\\
\texttt{authorized to perform this operation.}\\
\texttt{ErrorCode: AuthorizationFailure}\\
\texttt{\textless?xml version="1.0" encoding="utf-8"?\textgreater\textless Error\textgreater\textless Code\textgreater AuthorizationFailure\textless /Code\textgreater}\\
\texttt{\textless Message\textgreater This request is not authorized to perform this operation.}\\
\texttt{RequestId:c72fbeda-501e-0089-3c9f-e78298000000}\\
\texttt{Time:2019-03-31T08:58:06.2059267Z\textless /Message\textgreater\textless /Error\textgreater.}

\textcolor[HTML]{CC0000}{Error}: failed to pull data from the cloud - This request is not authorized to perform this operation. ErrorCode: AuthorizationFailure \\
\texttt{\textless?xml version="1.0" encoding="utf-8"?\textgreater\textless Error\textgreater\textless Code\textgreater
AuthorizationFailure}\\
\texttt{\textless /Code\textgreater\textless Message\textgreater This request is not authorized to perform this operation. \\
RequestId:c72fbeda-501e-0089-3c9f-e78298000000 \\
Time:2019-03-31T08:58:06.2059267Z\textless /Message\textgreater\textless /Error\textgreater}

\textcolor[HTML]{C4A000}{Having any troubles?} Hit us up at \href{https://dvc.org/support}{https://dvc.org/support}, we are always happy to help!

Please provide information about your setup

DVC version: 0.29.0+220b4b.mod

linux x86\_64 py3.6 pip\\

\vspace{0em}

\textbf{\#\#\# Issue 1515:}\\
\texttt{\$ dvc -V \\
0.23.2+bad2ef.mod}

DVC creates hardlinks instead of reflinks in APFS.

File system:

\texttt{\$ diskutil info / | grep -i system \\
File System Personality:   APFS}

\end{problemstatementboxA}
\clearpage
\begin{problemstatementboxA}[title=Problem Statement: iterative\_\_dvc\_0.33.1\_0.34.0 (continued)]

See last two lines:

\texttt{\$ dvc add Tags.xml -v \\
Debug: PRAGMA user\_version; \\
Debug: fetched: [(3,)] \\
Debug: CREATE TABLE IF NOT EXISTS state (inode INTEGER PRIMARY KEY, mtime TEXT NOT NULL, size TEXT NOT NULL, md5 TEXT NOT NULL, timestamp TEXT NOT NULL) \\
Debug: CREATE TABLE IF NOT EXISTS state\_info (count INTEGER) \\
Debug: CREATE TABLE IF NOT EXISTS link\_state (path TEXT PRIMARY KEY, inode INTEGER NOT NULL, mtime TEXT NOT NULL) \\
Debug: INSERT OR IGNORE INTO state\_info (count) SELECT 0 WHERE NOT EXISTS (SELECT * FROM state\_info) \\
Debug: PRAGMA user\_version = 3; \\
Debug: Skipping copying for '/Users/dmitry/src/modules-example/tmp/\\
test/Tags.xml', since it is not a symlink or a hardlink. \\
Adding 'Tags.xml' to '.gitignore'. \\
Saving 'Tags.xml' to cache '.dvc/cache'. \\
Debug: Path /Users/dmitry/src/modules-example/tmp/test/Tags.xml inode 12888021464 \\
Debug: SELECT * from state WHERE inode=12888021464 \\
Debug: fetched: [] \\
Debug: INSERT INTO state(inode, mtime, size, md5, timestamp) VALUES (12888021464, "1547854167848187904", "0", "6d861a1605e60b2f3a977a5a2a4419a2", "1547854178609126912") \\
Debug: File '/Users/dmitry/src/modules-example/tmp/test/.dvc/cache/6d/\\
861a1605e60b2f3a977a5a2a4419a2', md5 '6d861a1605e60b2f3a977a5a2a4419a2', actual 'None' \\
Debug: Cache type 'reflink' is not supported: reflink is not supported \\
Debug: Created 'hardlink': /Users/dmitry/src/modules-example/tmp/\\
test/.dvc/cache/6d/861a1605e60b2f3a977a5a2a4419a2 \textrightarrow  /Users/dmitry/src/\\
modules-example/tmp/test/Tags.xml}
\end{problemstatementboxA}
\vspace{1ex}

\begin{problemstatementboxB}
    \label{box:ex2}
    \textbf{2023.7.0}\\
Released on July 7, 2023

\textbf{Enhancements}
\begin{itemize}[leftmargin=*]
    \item Catch exceptions when attempting to load CLI entry points (\#10380) Jacob Tomlinson
\end{itemize}

\textbf{Bug Fixes}
\begin{itemize}[leftmargin=*]
    \item Fix typo in \texttt{\_clean\_ipython\_traceback} (\#10385) Alexander Clausen
    \item Ensure that \texttt{df} is immutable after \texttt{from\_pandas} (\#10383) Patrick Hoefler
    \item Warn consistently for \texttt{inplace} in \texttt{Series.rename} (\#10313) Patrick Hoefler
\end{itemize}

\textbf{Documentation}
\begin{itemize}[leftmargin=*]
    \item Add clarification about output shape and reshaping in rechunk documentation (\#10377) Swayam Patil
\end{itemize}

\textbf{Maintenance}
\begin{itemize}[leftmargin=*]
    \item Simplify \texttt{astype} implementation (\#10393) Patrick Hoefler
    \item Fix \texttt{test\_first\_and\_last} to accommodate deprecated \texttt{last} (\#10373) James Bourbeau
    \item Add \texttt{level} to \texttt{create\_merge\_tree} (\#10391) Patrick Hoefler
    \item Do not derive from \texttt{scipy.stats.chisquare} docstring (\#10382) Doug Davis
\end{itemize}

\vspace{0em}

\textbf{\#\#\# PR 10385:}\\
Regression from (\#10354) \href{https://github.com/dask/dask/pull/10354}{https://api.github.com/repos/dask/dask/pull/10354} - exception types of unhandled exceptions in ipython contexts were mangled and always set to type.

To reproduce, run the following in a jupyter notebook cell with dask 2023.6.1:

\texttt{import dask \\
raise TypeError("wat")}

Then, observe the websocket connection to jupyter; the message with \texttt{msg\_type="error"} looks like this:

\href{https://private-user-images.githubusercontent.com/5778/249846549-088f3a13-de7f-4430-a01a-63fc9e6ac3ec.png?jwt=eyJ0eXAiOiJKV1QiLCJhbGciOiJIUzI1NiJ9.eyJpc3MiOiJnaXRodWIuY29tIiwiYXVkIjoicmF3LmdpdGh1YnVzZXJjb250ZW50LmNvbSIsImtleSI6ImtleTUiLCJleHAiOjE3NjE2NzI3NzAsIm5iZiI6MTc2MTY3MjQ3MCwicGF0aCI6Ii81Nzc4LzI0OTg0NjU0OS0wODhmM2ExMy1kZTdmLTQ0MzAtYTAxYS02M2ZjOWU2YWMzZWMucG5nP1gtQW16LUFsZ29yaXRobT1BV1M0LUhNQUMtU0hBMjU2JlgtQW16LUNyZWRlbnRpYWw9QUtJQVZDT0RZTFNBNTNQUUs0WkElMkYyMDI1MTAyOCUyRnVzLWVhc3QtMSUyRnMzJTJGYXdzNF9yZXF1ZXN0JlgtQW16LURhdGU9MjAyNTEwMjhUMTcyNzUwWiZYLUFtei1FeHBpcmVzPTMwMCZYLUFtei1TaWduYXR1cmU9Yzg2NDQ5MGY3YjA3ZjM5ZWUxZjMwYjkwYTMyNTZiODM5NWNhOGM1MGNiYmQ2YmVkMzhjMmYzOWEzZGViYzlhMyZYLUFtei1TaWduZWRIZWFkZXJzPWhvc3QifQ.FzLVStB5-4M80q0jRDq0ge3a6WgyOxCAkJ4TWGoVsRY}{(Image Link)}

Notice \texttt{content["ename"]} is set to "type".

With dask 2023.6.0, \texttt{content["ename"]} is properly set to "TypeError":

\href{https://private-user-images.githubusercontent.com/5778/249847246-095b740b-5a54-4ebd-8ae4-781bcd143b45.png?jwt=eyJ0eXAiOiJKV1QiLCJhbGciOiJIUzI1NiJ9.eyJpc3MiOiJnaXRodWIuY29tIiwiYXVkIjoicmF3LmdpdGh1YnVzZXJjb250ZW50LmNvbSIsImtleSI6ImtleTUiLCJleHAiOjE3NjE2NzI3NzAsIm5iZiI6MTc2MTY3MjQ3MCwicGF0aCI6Ii81Nzc4LzI0OTg0NzI0Ni0wOTViNzQwYi01YTU0LTRlYmQtOGFlNC03ODFiY2QxNDNiNDUucG5nP1gtQW16LUFsZ29yaXRobT1BV1M0LUhNQUMtU0hBMjU2JlgtQW16LUNyZWRlbnRpYWw9QUtJQVZDT0RZTFNBNTNQUUs0WkElMkYyMDI1MTAyOCUyRnVzLWVhc3QtMSUyRnMzJTJGYXdzNF9yZXF1ZXN0JlgtQW16LURhdGU9MjAyNTEwMjhUMTcyNzUwWiZYLUFtei1FeHBpcmVzPTMwMCZYLUFtei1TaWduYXR1cmU9N2UzYTczNmQyMzIyZTM4ZDA2NjlkNTAzOTFlYjNiYWQxNTFjMTE3YTVlN2E1ZmNmYWIwMjBhZmU4NDYzMDk5ZSZYLUFtei1TaWduZWRIZWFkZXJzPWhvc3QifQ._udG---XFgq4pMSJNyT7rhq0wew9kN-0ToAt3_5SnTc}{(Image Link)}

These values aren't inconsequential - they end up being serialized into the ipynb file, and cause hard to debug issues.

(nb: I would prefer if this kind of issue was not something that can be caused by \texttt{import dask}, by opting in to this functionality by configuration or actively registering the hook functions)

\begin{itemize}[leftmargin=*]
    \item Tests added / passed
    \item Passes \texttt{pre-commit run -{}-all-files}
\end{itemize}

Thanks to \href{https://github.com/matbryan52}{(@matbryan52)} for the help debugging this.

\end{problemstatementboxB}
\clearpage
\begin{problemstatementboxB}[title=Problem Statement: dask\_\_dask\_2023.6.1\_2023.7.0 (continued)]

\textbf{\#\#\# PR 10383:}
\begin{itemize}[leftmargin=*]
    \item Tests added / passed
    \item Passes \texttt{pre-commit run -{}-all-files}
\end{itemize}
We had a similar problem in dask-expr\\

\vspace{0em}

\textbf{\#\#\# PR 10393:}
\begin{itemize}[leftmargin=*]
    \item Tests added / passed
    \item Passes \texttt{pre-commit run -{}-all-files}
\end{itemize}

\vspace{0em}

\textbf{\#\#\# PR 10391:}
\begin{itemize}[leftmargin=*]
    \item Passes \texttt{pre-commit run -{}-all-files}
\end{itemize}

\vspace{0em}

\textbf{\#\#\# PR 10373:}\\
This fixes

\texttt{FAILED dask/dataframe/tests/test\_dataframe.py::}\\
\texttt{test\_first\_and\_last[last] - FutureWarning: last is deprecated}\\
\texttt{and will be removed in a future version. Please create a mask}\\
\texttt{and filter using \`.loc\` instead}

which we're currently seeing in the upstream build (xref (\#10347) \href{https://github.com/dask/dask/actions/runs/6562188664}{(Workflow Run URL)}

Python 3.10 Test Summary \\
\texttt{dask/dataframe/tests/}\\
\texttt{test\_multi.py::test\_concat\_categorical[True-False-True]:}\\
\texttt{[XPASS(strict)] fails on pandas dev:}\\
\texttt{\href{https://github.com/dask/dask/issues/10558}{https://api.github.com/repos/dask/dask/issues/10558}}\\
\texttt{dask/dataframe/tests/}\\
\texttt{test\_multi.py::test\_concat\_categorical[False-False-True]:}\\
\texttt{[XPASS(strict)] fails on pandas dev:}\\
\texttt{\href{https://github.com/dask/dask/issues/10558}{https://api.github.com/repos/dask/dask/issues/10558}}

Note that we were already emitting a FutureWarning for \texttt{DataFrame.last}, so this is a tests-only PR

cc \href{https://github.com/j-bennet}{(@j-bennet)} \href{https://github.com/phofl}{(@phofl)}\\

\vspace{0em}

\textbf{\#\#\# PR 10382:}
\begin{itemize}[leftmargin=*]
    \item Closes (\#10381) Looks like our docs are running into an error on main. See this build from a recent PR \href{https://readthedocs.org/projects/dask/builds/21146384/}{https://readthedocs.org/projects/dask/builds/21146384/}
    \item [x] Tests added / passed
    \item Passes \texttt{pre-commit run -{}-all-files}
\end{itemize}

\texttt{scipy.stats.chisquare} recently started using a doi Sphinx directive that exists in their repo but it's not importable. This is breaking dask docs generation. We can just point folks to that docstring instead of deriving from it.

for reference here is the PR in scipy: \href{https://github.com/scipy/scipy/pull/17682}{(scipy/scipy\#17682)}\\

\vspace{0em}

\textbf{\#\#\# PR 10377:}
\begin{itemize}[leftmargin=*]
    \item [] Closes (\#10361) It is not documented anywhere that \texttt{dask.array.rechunk} expects the output to have the same shape as the input and does not allow reshaping. The validation step is also quite hidden. This has led to some confusion (\href{https://github.com/dask/distributed/pull/7897\#discussion_r1234169191}{dask/distributed\#7897 (comment)}), so I think it would be good to add a remark about this in the documentation.
    \item Tests added / passed
    \item Passes \texttt{pre-commit run -{}-all-files}
\end{itemize}

\vspace{0em}

\textbf{\#\#\# PR 10313:}\\
The previous warning seemed very inconsistent\\

\vspace{0em}

\textbf{\#\#\# PR 10380:}
\begin{itemize}[leftmargin=*]
    \item Closes (\#10379) If a package registers an entry point for the CLI but the entrypoint is broken or not importable the CLI cannot work at all.
\end{itemize}

This is a little hard to create an MRE for but I seeing this with dask-ctl when updating textual to a more recent version. The error comes from the following line.

\href{https://github.com/dask/dask/blob/85c99bc20abc382774cfb6e5bf5f2db76ac09378/dask/cli.py\#L97}{(dask/dask/cli.py)}

Line 97 in \path{/dask/dask/commit/85c99bc20abc382774cfb6e5bf5f2db76ac09378}

\texttt{command = entry\_point.load()}

When this line is called third-party code is loaded, so we have no control over what happens. We could broadly catch exceptions here, print a warning and carry on.

\begin{itemize}[leftmargin=*]
    \item Tests added / passed
    \item Passes \texttt{pre-commit run -{}-all-files}
\end{itemize}
\end{problemstatementboxB}
\clearpage

%% file: checklist.tex
\section*{NeurIPS Paper Checklist}

\begin{enumerate}

\item {\bf Claims}
    \item[] Question: Do the main claims made in the abstract and introduction accurately reflect the paper's contributions and scope?
    \item[] Answer: \answerYes{}
    \item[] Justification: The abstract and introduction state that \tool is a benchmark for long-horizon software evolution, describe the construction scope, and summarize the experimental findings. The claims are supported by the benchmark construction in Section~\ref{sec:dataset} and the results in Section~\ref{sec:experiments}.
    \item[] Guidelines:
    \begin{itemize}
        \item The answer \answerNA{} means that the abstract and introduction do not include the claims made in the paper.
        \item The abstract and/or introduction should clearly state the claims made, including the contributions made in the paper and important assumptions and limitations. A \answerNo{} or \answerNA{} answer to this question will not be perceived well by the reviewers. 
        \item The claims made should match theoretical and experimental results, and reflect how much the results can be expected to generalize to other settings. 
        \item It is fine to include aspirational goals as motivation as long as it is clear that these goals are not attained by the paper. 
    \end{itemize}

\item {\bf Limitations}
    \item[] Question: Does the paper discuss the limitations of the work performed by the authors?
    \item[] Answer: \answerYes{}
    \item[] Justification: The paper includes a dedicated Limitations section after the conclusion, covering language coverage, release-note specifications, benchmark scale, contamination risk, and Fix Rate granularity.
    \item[] Guidelines:
    \begin{itemize}
        \item The answer \answerNA{} means that the paper has no limitation while the answer \answerNo{} means that the paper has limitations, but those are not discussed in the paper. 
        \item The authors are encouraged to create a separate ``Limitations'' section in their paper.
        \item The paper should point out any strong assumptions and how robust the results are to violations of these assumptions (e.g., independence assumptions, noiseless settings, model well-specification, asymptotic approximations only holding locally). The authors should reflect on how these assumptions might be violated in practice and what the implications would be.
        \item The authors should reflect on the scope of the claims made, e.g., if the approach was only tested on a few datasets or with a few runs. In general, empirical results often depend on implicit assumptions, which should be articulated.
        \item The authors should reflect on the factors that influence the performance of the approach. For example, a facial recognition algorithm may perform poorly when image resolution is low or images are taken in low lighting. Or a speech-to-text system might not be used reliably to provide closed captions for online lectures because it fails to handle technical jargon.
        \item The authors should discuss the computational efficiency of the proposed algorithms and how they scale with dataset size.
        \item If applicable, the authors should discuss possible limitations of their approach to address problems of privacy and fairness.
        \item While the authors might fear that complete honesty about limitations might be used by reviewers as grounds for rejection, a worse outcome might be that reviewers discover limitations that aren't acknowledged in the paper. The authors should use their best judgment and recognize that individual actions in favor of transparency play an important role in developing norms that preserve the integrity of the community. Reviewers will be specifically instructed to not penalize honesty concerning limitations.
    \end{itemize}

\item {\bf Theory assumptions and proofs}
    \item[] Question: For each theoretical result, does the paper provide the full set of assumptions and a complete (and correct) proof?
    \item[] Answer: \answerNA{}
    \item[] Justification: The paper introduces a benchmark and empirical evaluation metric, but does not present theoretical results or proofs.
    \item[] Guidelines:
    \begin{itemize}
        \item The answer \answerNA{} means that the paper does not include theoretical results. 
        \item All the theorems, formulas, and proofs in the paper should be numbered and cross-referenced.
        \item All assumptions should be clearly stated or referenced in the statement of any theorems.
        \item The proofs can either appear in the main paper or the supplemental material, but if they appear in the supplemental material, the authors are encouraged to provide a short proof sketch to provide intuition. 
        \item Inversely, any informal proof provided in the core of the paper should be complemented by formal proofs provided in appendix or supplemental material.
        \item Theorems and Lemmas that the proof relies upon should be properly referenced. 
    \end{itemize}

    \item {\bf Experimental result reproducibility}
    \item[] Question: Does the paper fully disclose all the information needed to reproduce the main experimental results of the paper to the extent that it affects the main claims and/or conclusions of the paper (regardless of whether the code and data are provided or not)?
    \item[] Answer: \answerYes{}
    \item[] Justification: Sections~\ref{sec:dataset} and~\ref{sec:experiments} describe task construction, input format, evaluation metrics, scaffolds, and model settings, with additional details and examples in the appendix.
    \item[] Guidelines:
    \begin{itemize}
        \item The answer \answerNA{} means that the paper does not include experiments.
        \item If the paper includes experiments, a \answerNo{} answer to this question will not be perceived well by the reviewers: Making the paper reproducible is important, regardless of whether the code and data are provided or not.
        \item If the contribution is a dataset and\slash or model, the authors should describe the steps taken to make their results reproducible or verifiable. 
        \item Depending on the contribution, reproducibility can be accomplished in various ways. For example, if the contribution is a novel architecture, describing the architecture fully might suffice, or if the contribution is a specific model and empirical evaluation, it may be necessary to either make it possible for others to replicate the model with the same dataset, or provide access to the model. In general. releasing code and data is often one good way to accomplish this, but reproducibility can also be provided via detailed instructions for how to replicate the results, access to a hosted model (e.g., in the case of a large language model), releasing of a model checkpoint, or other means that are appropriate to the research performed.
        \item While NeurIPS does not require releasing code, the conference does require all submissions to provide some reasonable avenue for reproducibility, which may depend on the nature of the contribution. For example
        \begin{enumerate}
            \item If the contribution is primarily a new algorithm, the paper should make it clear how to reproduce that algorithm.
            \item If the contribution is primarily a new model architecture, the paper should describe the architecture clearly and fully.
            \item If the contribution is a new model (e.g., a large language model), then there should either be a way to access this model for reproducing the results or a way to reproduce the model (e.g., with an open-source dataset or instructions for how to construct the dataset).
            \item We recognize that reproducibility may be tricky in some cases, in which case authors are welcome to describe the particular way they provide for reproducibility. In the case of closed-source models, it may be that access to the model is limited in some way (e.g., to registered users), but it should be possible for other researchers to have some path to reproducing or verifying the results.
        \end{enumerate}
    \end{itemize}

\item {\bf Open access to data and code}
    \item[] Question: Does the paper provide open access to the data and code, with sufficient instructions to faithfully reproduce the main experimental results, as described in supplemental material?
    \item[] Answer: \answerNo{}
    \item[] Justification: The current paper describes the planned release of task metadata, construction scripts, Docker-based execution configuration, evaluation harness, generated patches, and aggregate result files, but the anonymized release package or URL is not yet included in this submission bundle.
    \item[] Guidelines:
    \begin{itemize}
        \item The answer \answerNA{} means that paper does not include experiments requiring code.
        \item Please see the NeurIPS code and data submission guidelines (\url{https://neurips.cc/public/guides/CodeSubmissionPolicy}) for more details.
        \item While we encourage the release of code and data, we understand that this might not be possible, so \answerNo{} is an acceptable answer. Papers cannot be rejected simply for not including code, unless this is central to the contribution (e.g., for a new open-source benchmark).
        \item The instructions should contain the exact command and environment needed to run to reproduce the results. See the NeurIPS code and data submission guidelines (\url{https://neurips.cc/public/guides/CodeSubmissionPolicy}) for more details.
        \item The authors should provide instructions on data access and preparation, including how to access the raw data, preprocessed data, intermediate data, and generated data, etc.
        \item The authors should provide scripts to reproduce all experimental results for the new proposed method and baselines. If only a subset of experiments are reproducible, they should state which ones are omitted from the script and why.
        \item At submission time, to preserve anonymity, the authors should release anonymized versions (if applicable).
        \item Providing as much information as possible in supplemental material (appended to the paper) is recommended, but including URLs to data and code is permitted.
    \end{itemize}

\item {\bf Experimental setting/details}
    \item[] Question: Does the paper specify all the training and test details (e.g., data splits, hyperparameters, how they were chosen, type of optimizer) necessary to understand the results?
    \item[] Answer: \answerYes{}
    \item[] Justification: The experimental section specifies the agent scaffolds, model set, benchmark setting, and metrics, while the appendix provides additional task fields, context construction, and problem-statement examples.
    \item[] Guidelines:
    \begin{itemize}
        \item The answer \answerNA{} means that the paper does not include experiments.
        \item The experimental setting should be presented in the core of the paper to a level of detail that is necessary to appreciate the results and make sense of them.
        \item The full details can be provided either with the code, in appendix, or as supplemental material.
    \end{itemize}

\item {\bf Experiment statistical significance}
    \item[] Question: Does the paper report error bars suitably and correctly defined or other appropriate information about the statistical significance of the experiments?
    \item[] Answer: \answerNo{}
    \item[] Justification: The paper reports descriptive benchmark results and does not make statistical-significance claims.
    \item[] Guidelines:
    \begin{itemize}
        \item The answer \answerNA{} means that the paper does not include experiments.
        \item The authors should answer \answerYes{} if the results are accompanied by error bars, confidence intervals, or statistical significance tests, at least for the experiments that support the main claims of the paper.
        \item The factors of variability that the error bars are capturing should be clearly stated (for example, train/test split, initialization, random drawing of some parameter, or overall run with given experimental conditions).
        \item The method for calculating the error bars should be explained (closed form formula, call to a library function, bootstrap, etc.)
        \item The assumptions made should be given (e.g., Normally distributed errors).
        \item It should be clear whether the error bar is the standard deviation or the standard error of the mean.
        \item It is OK to report 1-sigma error bars, but one should state it. The authors should preferably report a 2-sigma error bar than state that they have a 96\% CI, if the hypothesis of Normality of errors is not verified.
        \item For asymmetric distributions, the authors should be careful not to show in tables or figures symmetric error bars that would yield results that are out of range (e.g., negative error rates).
        \item If error bars are reported in tables or plots, the authors should explain in the text how they were calculated and reference the corresponding figures or tables in the text.
    \end{itemize}

\item {\bf Experiments compute resources}
    \item[] Question: For each experiment, does the paper provide sufficient information on the computer resources (type of compute workers, memory, time of execution) needed to reproduce the experiments?
    \item[] Answer: \answerNo{}
    \item[] Justification: The paper describes models and scaffolds, but not detailed compute-resource accounting.
    \item[] Guidelines:
    \begin{itemize}
        \item The answer \answerNA{} means that the paper does not include experiments.
        \item The paper should indicate the type of compute workers CPU or GPU, internal cluster, or cloud provider, including relevant memory and storage.
        \item The paper should provide the amount of compute required for each of the individual experimental runs as well as estimate the total compute. 
        \item The paper should disclose whether the full research project required more compute than the experiments reported in the paper (e.g., preliminary or failed experiments that didn't make it into the paper). 
    \end{itemize}
    
\item {\bf Code of ethics}
    \item[] Question: Does the research conducted in the paper conform, in every respect, with the NeurIPS Code of Ethics \url{https://neurips.cc/public/EthicsGuidelines}?
    \item[] Answer: \answerYes{}
    \item[] Justification: The work evaluates coding agents on public software repositories and does not involve human-subject experiments, private data, or deployment affecting individuals.
    \item[] Guidelines:
    \begin{itemize}
        \item The answer \answerNA{} means that the authors have not reviewed the NeurIPS Code of Ethics.
        \item If the authors answer \answerNo, they should explain the special circumstances that require a deviation from the Code of Ethics.
        \item The authors should make sure to preserve anonymity (e.g., if there is a special consideration due to laws or regulations in their jurisdiction).
    \end{itemize}

\item {\bf Broader impacts}
    \item[] Question: Does the paper discuss both potential positive societal impacts and negative societal impacts of the work performed?
    \item[] Answer: \answerYes{}
    \item[] Justification: Positive impacts include more rigorous evaluation of coding agents for software maintenance and safer benchmarking of long-horizon engineering tasks. Potential negative impacts include accelerating capability development for agents that could be misused for automated vulnerability discovery, malicious code authoring, or large-scale repository abuse; the paper's focus on public repositories, benchmark evaluation, and release-level auditing partially mitigates these risks.
    \item[] Guidelines:
    \begin{itemize}
        \item The answer \answerNA{} means that there is no societal impact of the work performed.
        \item If the authors answer \answerNA{} or \answerNo, they should explain why their work has no societal impact or why the paper does not address societal impact.
        \item Examples of negative societal impacts include potential malicious or unintended uses (e.g., disinformation, generating fake profiles, surveillance), fairness considerations (e.g., deployment of technologies that could make decisions that unfairly impact specific groups), privacy considerations, and security considerations.
        \item The conference expects that many papers will be foundational research and not tied to particular applications, let alone deployments. However, if there is a direct path to any negative applications, the authors should point it out. For example, it is legitimate to point out that an improvement in the quality of generative models could be used to generate Deepfakes for disinformation. On the other hand, it is not needed to point out that a generic algorithm for optimizing neural networks could enable people to train models that generate Deepfakes faster.
        \item The authors should consider possible harms that could arise when the technology is being used as intended and functioning correctly, harms that could arise when the technology is being used as intended but gives incorrect results, and harms following from (intentional or unintentional) misuse of the technology.
        \item If there are negative societal impacts, the authors could also discuss possible mitigation strategies (e.g., gated release of models, providing defenses in addition to attacks, mechanisms for monitoring misuse, mechanisms to monitor how a system learns from feedback over time, improving the efficiency and accessibility of ML).
    \end{itemize}
    
\item {\bf Safeguards}
    \item[] Question: Does the paper describe safeguards that have been put in place for responsible release of data or models that have a high risk for misuse (e.g., pre-trained language models, image generators, or scraped datasets)?
    \item[] Answer: \answerNA{}
    \item[] Justification: The paper does not release high-risk models, scraped media, or user-facing systems. The released benchmark consists of task metadata and evaluation artifacts derived from public software repositories.
    \item[] Guidelines:
    \begin{itemize}
        \item The answer \answerNA{} means that the paper poses no such risks.
        \item Released models that have a high risk for misuse or dual-use should be released with necessary safeguards to allow for controlled use of the model, for example by requiring that users adhere to usage guidelines or restrictions to access the model or implementing safety filters. 
        \item Datasets that have been scraped from the Internet could pose safety risks. The authors should describe how they avoided releasing unsafe images.
        \item We recognize that providing effective safeguards is challenging, and many papers do not require this, but we encourage authors to take this into account and make a best faith effort.
    \end{itemize}

\item {\bf Licenses for existing assets}
    \item[] Question: Are the creators or original owners of assets (e.g., code, data, models), used in the paper, properly credited and are the license and terms of use explicitly mentioned and properly respected?
    \item[] Answer: \answerNo{}
    \item[] Justification: The paper credits SWE-bench, SWE-gym, the evaluated coding-agent scaffolds, and the evaluated model families, but it does not yet enumerate repository-level licenses and terms of use for all upstream software assets in the submission itself. The benchmark release should include that metadata explicitly.
    \item[] Guidelines:
    \begin{itemize}
        \item The answer \answerNA{} means that the paper does not use existing assets.
        \item The authors should cite the original paper that produced the code package or dataset.
        \item The authors should state which version of the asset is used and, if possible, include a URL.
        \item The name of the license (e.g., CC-BY 4.0) should be included for each asset.
        \item For scraped data from a particular source (e.g., website), the copyright and terms of service of that source should be provided.
        \item If assets are released, the license, copyright information, and terms of use in the package should be provided. For popular datasets, \url{paperswithcode.com/datasets} has curated licenses for some datasets. Their licensing guide can help determine the license of a dataset.
        \item For existing datasets that are re-packaged, both the original license and the license of the derived asset (if it has changed) should be provided.
        \item If this information is not available online, the authors are encouraged to reach out to the asset's creators.
    \end{itemize}

\item {\bf New assets}
    \item[] Question: Are new assets introduced in the paper well documented and is the documentation provided alongside the assets?
    \item[] Answer: \answerYes{}
    \item[] Justification: \tool is a new benchmark asset, and the paper documents its construction, task fields, evaluation metrics, and problem-statement format in Section~\ref{sec:dataset} and the appendix.
    \item[] Guidelines:
    \begin{itemize}
        \item The answer \answerNA{} means that the paper does not release new assets.
        \item Researchers should communicate the details of the dataset\slash code\slash model as part of their submissions via structured templates. This includes details about training, license, limitations, etc. 
        \item The paper should discuss whether and how consent was obtained from people whose asset is used.
        \item At submission time, remember to anonymize your assets (if applicable). You can either create an anonymized URL or include an anonymized zip file.
    \end{itemize}

\item {\bf Crowdsourcing and research with human subjects}
    \item[] Question: For crowdsourcing experiments and research with human subjects, does the paper include the full text of instructions given to participants and screenshots, if applicable, as well as details about compensation (if any)? 
    \item[] Answer: \answerNA{}
    \item[] Justification: The paper does not involve crowdsourcing or research with human subjects.
    \item[] Guidelines:
    \begin{itemize}
        \item The answer \answerNA{} means that the paper does not involve crowdsourcing nor research with human subjects.
        \item Including this information in the supplemental material is fine, but if the main contribution of the paper involves human subjects, then as much detail as possible should be included in the main paper. 
        \item According to the NeurIPS Code of Ethics, workers involved in data collection, curation, or other labor should be paid at least the minimum wage in the country of the data collector. 
    \end{itemize}

\item {\bf Institutional review board (IRB) approvals or equivalent for research with human subjects}
    \item[] Question: Does the paper describe potential risks incurred by study participants, whether such risks were disclosed to the subjects, and whether Institutional Review Board (IRB) approvals (or an equivalent approval/review based on the requirements of your country or institution) were obtained?
    \item[] Answer: \answerNA{}
    \item[] Justification: The paper does not involve crowdsourcing or human-subject research and therefore does not require IRB approval.
    \item[] Guidelines:
    \begin{itemize}
        \item The answer \answerNA{} means that the paper does not involve crowdsourcing nor research with human subjects.
        \item Depending on the country in which research is conducted, IRB approval (or equivalent) may be required for any human subjects research. If you obtained IRB approval, you should clearly state this in the paper. 
        \item We recognize that the procedures for this may vary significantly between institutions and locations, and we expect authors to adhere to the NeurIPS Code of Ethics and the guidelines for their institution. 
        \item For initial submissions, do not include any information that would break anonymity (if applicable), such as the institution conducting the review.
    \end{itemize}

\item {\bf Declaration of LLM usage}
    \item[] Question: Does the paper describe the usage of LLMs if it is an important, original, or non-standard component of the core methods in this research? Note that if the LLM is used only for writing, editing, or formatting purposes and does \emph{not} impact the core methodology, scientific rigor, or originality of the research, declaration is not required.
    \item[] Answer: \answerYes{}
    \item[] Justification: The paper describes the evaluated LLMs and coding-agent settings.
    \item[] Guidelines:
    \begin{itemize}
        \item The answer \answerNA{} means that the core method development in this research does not involve LLMs as any important, original, or non-standard components.
        \item Please refer to our LLM policy in the NeurIPS handbook for what should or should not be described.
    \end{itemize}

\end{enumerate}

%% file: main.bbl
\begin{thebibliography}{89}
\providecommand{\natexlab}[1]{#1}
\providecommand{\url}[1]{\texttt{#1}}
\expandafter\ifx\csname urlstyle\endcsname\relax
  \providecommand{\doi}[1]{doi: #1}\else
  \providecommand{\doi}{doi: \begingroup \urlstyle{rm}\Url}\fi

\bibitem[Asai et~al.(2024)Asai, Wu, Wang, Sil, and Hajishirzi]{asai2023self}
Akari Asai, Zeqiu Wu, Yizhong Wang, Avirup Sil, and Hannaneh Hajishirzi.
\newblock Self-rag: Learning to retrieve, generate, and critique through
  self-reflection.
\newblock \emph{International Conference on Learning Representations}, 2024.

\bibitem[Austin et~al.(2021)Austin, Odena, Nye, Bosma, Michalewski, Dohan,
  Jiang, Cai, Terry, Le, et~al.]{austin2021program}
Jacob Austin, Augustus Odena, Maxwell Nye, Maarten Bosma, Henryk Michalewski,
  David Dohan, Ellen Jiang, Carrie Cai, Michael Terry, Quoc Le, et~al.
\newblock Program synthesis with large language models.
\newblock \emph{arXiv preprint arXiv:2108.07732}, 2021.

\bibitem[Badertdinov et~al.(2025)Badertdinov, Golubev, Nekrashevich, Shevtsov,
  Karasik, Andriushchenko, Trofimova, Litvintseva, and
  Yangel]{badertdinov2025swe}
Ibragim Badertdinov, Alexander Golubev, Maksim Nekrashevich, Anton Shevtsov,
  Simon Karasik, Andrei Andriushchenko, Maria Trofimova, Daria Litvintseva, and
  Boris Yangel.
\newblock Swe-rebench: An automated pipeline for task collection and
  decontaminated evaluation of software engineering agents.
\newblock \emph{arXiv preprint arXiv:2505.20411}, 2025.

\bibitem[Bui et~al.(2023)Bui, Le, Wang, Li, Gotmare, and Hoi]{bui2023codetf}
Nghi~DQ Bui, Hung Le, Yue Wang, Junnan Li, Akhilesh~Deepak Gotmare, and
  Steven~CH Hoi.
\newblock Codetf: One-stop transformer library for state-of-the-art code llm.
\newblock \emph{arXiv preprint arXiv:2306.00029}, 2023.

\bibitem[Cai et~al.(2024)Cai, Wang, Ma, Chen, and Zhou]{llmastool}
Tianle Cai, Xuezhi Wang, Tengyu Ma, Xinyun Chen, and Denny Zhou.
\newblock Large language models as tool makers.
\newblock In \emph{International Conference on Representation Learning}, 2024.

\bibitem[Carbonneaux et~al.(2025)Carbonneaux, Cohen, Gehring, Kahn, Kossen,
  Kreuk, McMilin, Meyer, Wei, Zhang, et~al.]{carbonneaux2025cwm}
Quentin Carbonneaux, Gal Cohen, Jonas Gehring, Jacob Kahn, Jannik Kossen, Felix
  Kreuk, Emily McMilin, Michel Meyer, Yuxiang Wei, David Zhang, et~al.
\newblock Cwm: An open-weights llm for research on code generation with world
  models.
\newblock \emph{arXiv preprint arXiv:2510.02387}, 2025.

\bibitem[Chen et~al.(2025)Chen, Li, Gong, Jiang, Fei, Yang, Shan, Yu, Wang,
  Zhu, et~al.]{chen2025minimax}
Aili Chen, Aonian Li, Bangwei Gong, Binyang Jiang, Bo~Fei, Bo~Yang, Boji Shan,
  Changqing Yu, Chao Wang, Cheng Zhu, et~al.
\newblock Minimax-m1: Scaling test-time compute efficiently with lightning
  attention.
\newblock \emph{arXiv preprint arXiv:2506.13585}, 2025.

\bibitem[Chen et~al.(2022)Chen, Zhang, Nguyen, Zan, Lin, Lou, and
  Chen]{chen2022codet}
Bei Chen, Fengji Zhang, Anh Nguyen, Daoguang Zan, Zeqi Lin, Jian-Guang Lou, and
  Weizhu Chen.
\newblock Codet: Code generation with generated tests.
\newblock \emph{arXiv preprint arXiv:2207.10397}, 2022.

\bibitem[Chen et~al.(2021{\natexlab{a}})Chen, Tworek, Jun, Yuan, Pinto, Kaplan,
  Edwards, Burda, Joseph, Brockman, et~al.]{chen2021evaluating}
Mark Chen, Jerry Tworek, Heewoo Jun, Qiming Yuan, Henrique Ponde De~Oliveira
  Pinto, Jared Kaplan, Harri Edwards, Yuri Burda, Nicholas Joseph, Greg
  Brockman, et~al.
\newblock Evaluating large language models trained on code.
\newblock \emph{arXiv preprint arXiv:2107.03374}, 2021{\natexlab{a}}.

\bibitem[Chen et~al.(2021{\natexlab{b}})Chen, Tworek, Jun, Yuan, Pinto, Kaplan,
  Edwards, Burda, Joseph, Brockman, et~al.]{humaneval}
Mark Chen, Jerry Tworek, Heewoo Jun, Qiming Yuan, Henrique Ponde De~Oliveira
  Pinto, Jared Kaplan, Harri Edwards, Yuri Burda, Nicholas Joseph, Greg
  Brockman, et~al.
\newblock Evaluating large language models trained on code.
\newblock \emph{arXiv preprint arXiv:2107.03374}, 2021{\natexlab{b}}.

\bibitem[Chen et~al.(2023)Chen, Lin, Sch{\"a}rli, and Zhou]{chen2023teaching}
Xinyun Chen, Maxwell Lin, Nathanael Sch{\"a}rli, and Denny Zhou.
\newblock Teaching large language models to self-debug.
\newblock \emph{arXiv preprint arXiv:2304.05128}, 2023.

\bibitem[{Cognition}(2024)]{devin}
{Cognition}.
\newblock Devin ai, 2024.
\newblock \url{https://cognition.ai/blog/introducing-devin}.

\bibitem[{DeepSeek-AI}(2025{\natexlab{a}})]{deepseek_r1_0528_2025}
{DeepSeek-AI}.
\newblock Deepseek-r1-0528 release.
\newblock \url{https://api-docs.deepseek.com/news/news250528},
  2025{\natexlab{a}}.
\newblock Accessed 2026-03-12.

\bibitem[{DeepSeek-AI}(2025{\natexlab{b}})]{deepseek_v31_2025}
{DeepSeek-AI}.
\newblock Deepseek-v3.1 release.
\newblock \url{https://api-docs.deepseek.com/news/news250821},
  2025{\natexlab{b}}.
\newblock Accessed 2026-03-12.

\bibitem[{DeepSeek-AI}(2025{\natexlab{c}})]{deepseek_v32_2025}
{DeepSeek-AI}.
\newblock Deepseek-v3.2: Pushing the frontier of open large language models.
\newblock \emph{arXiv preprint arXiv:2512.02556}, 2025{\natexlab{c}}.

\bibitem[{DeepSeek AI}(2025)]{deepseekv31}
{DeepSeek AI}.
\newblock {DeepSeek V3.1}, 2025.
\newblock \url{https://api-docs.deepseek.com/news/news250821}.

\bibitem[{DORA Research Program}(2025)]{dora2025ai}
{DORA Research Program}.
\newblock State of ai-assisted software development: 2025 dora report.
\newblock
  \url{https://cloud.google.com/resources/content/2025-dora-ai-assisted-software-development-report},
  September 2025.
\newblock DevOps Research and Assessment (DORA).

\bibitem[Fan et~al.(2023)Fan, Gokkaya, Harman, Lyubarskiy, Sengupta, Yoo, and
  Zhang]{fan2023large}
Angela Fan, Beliz Gokkaya, Mark Harman, Mitya Lyubarskiy, Shubho Sengupta, Shin
  Yoo, and Jie~M Zhang.
\newblock Large language models for software engineering: Survey and open
  problems.
\newblock In \emph{2023 IEEE/ACM International Conference on Software
  Engineering: Future of Software Engineering (ICSE-FoSE)}, pages 31--53. IEEE,
  2023.

\bibitem[Gao et~al.(2025)Gao, Hu, Gao, Xia, and Jin]{gao2025current}
Cuiyun Gao, Xing Hu, Shan Gao, Xin Xia, and Zhi Jin.
\newblock The current challenges of software engineering in the era of large
  language models.
\newblock \emph{ACM Transactions on Software Engineering and Methodology},
  34\penalty0 (5):\penalty0 1--30, 2025.

\bibitem[Ge et~al.(2024)Ge, Hu, Wang, Chen, and Wei]{ge2023context}
Tao Ge, Jing Hu, Xun Wang, Si-Qing Chen, and Furu Wei.
\newblock In-context autoencoder for context compression in a large language
  model.
\newblock \emph{International Conference on Learning Representations}, 2024.

\bibitem[{GLM-4.5 Team}(2025)]{glm45team_2025}
{GLM-4.5 Team}.
\newblock Glm-4.5: Agentic, reasoning, and coding (arc) foundation models.
\newblock \emph{arXiv preprint arXiv:2508.06471}, 2025.

\bibitem[Gutierrez et~al.(2024)Gutierrez, Shu, Gu, Yasunaga, and
  Su]{gutierrez2024hipporag}
Bernal~Jimenez Gutierrez, Yiheng Shu, Yu~Gu, Michihiro Yasunaga, and Yu~Su.
\newblock Hipporag: Neurobiologically inspired long-term memory for large
  language models.
\newblock \emph{Advances in Neural Information Processing Systems}, 37, 2024.

\bibitem[He et~al.(2025)He, Treude, and Lo]{he2025llm}
Junda He, Christoph Treude, and David Lo.
\newblock Llm-based multi-agent systems for software engineering: Literature
  review, vision, and the road ahead.
\newblock \emph{ACM Transactions on Software Engineering and Methodology},
  34\penalty0 (5):\penalty0 1--30, 2025.

\bibitem[Hendrycks et~al.(2021)Hendrycks, Basart, Kadavath, Mazeika, Arora,
  Guo, Burns, Puranik, He, Song, et~al.]{hendrycks2021measuring}
Dan Hendrycks, Steven Basart, Saurav Kadavath, Mantas Mazeika, Akul Arora,
  Ethan Guo, Collin Burns, Samir Puranik, Horace He, Dawn Song, et~al.
\newblock Measuring coding challenge competence with apps.
\newblock \emph{arXiv preprint arXiv:2105.09938}, 2021.

\bibitem[Hoang et~al.(2025)Hoang, Le-Anh, Le, and Bui]{hoang2025codewiki}
Anh~Nguyen Hoang, Minh Le-Anh, Bach Le, and Nghi~DQ Bui.
\newblock Codewiki: Evaluating ai's ability to generate holistic documentation
  for large-scale codebases.
\newblock \emph{arXiv preprint arXiv:2510.24428}, 2025.

\bibitem[Hua et~al.(2025)Hua, Ye, Fu, Xiao, Cai, Wu, Lin, Wang, and
  Liu]{hua2025context}
Qishuo Hua, Lyumanshan Ye, Dayuan Fu, Yang Xiao, Xiaojie Cai, Yunze Wu, Jifan
  Lin, Junfei Wang, and Pengfei Liu.
\newblock Context engineering 2.0: The context of context engineering.
\newblock \emph{arXiv preprint arXiv:2510.26493}, 2025.

\bibitem[Huang et~al.(2023)Huang, Zhang, Luck, Bu, Qing, and
  Cui]{huang2023agentcoder}
Dong Huang, Jie~M Zhang, Michael Luck, Qingwen Bu, Yuhao Qing, and Heming Cui.
\newblock Agentcoder: Multi-agent-based code generation with iterative testing
  and optimisation.
\newblock \emph{arXiv preprint arXiv:2312.13010}, 2023.

\bibitem[Jain et~al.(2024)Jain, Synnaeve, and Roziere]{jain2024testgeneval}
Kush Jain, Gabriel Synnaeve, and Baptiste Roziere.
\newblock Testgeneval: A real world unit test generation and test completion
  benchmark.
\newblock \emph{arXiv preprint arXiv:2410.00752}, 2024.

\bibitem[Jimenez et~al.(2023)Jimenez, Yang, Wettig, Yao, Pei, Press, and
  Narasimhan]{jimenez2023swe}
Carlos~E Jimenez, John Yang, Alexander Wettig, Shunyu Yao, Kexin Pei, Ofir
  Press, and Karthik Narasimhan.
\newblock Swe-bench: Can language models resolve real-world github issues?
\newblock \emph{arXiv preprint arXiv:2310.06770}, 2023.

\bibitem[Jimenez et~al.(2024)Jimenez, Yang, Wettig, Yao, Pei, Press, and
  Narasimhan]{jimenez2024swebench}
Carlos~E. Jimenez, John Yang, Alexander Wettig, Shunyu Yao, Kexin Pei, Ofir
  Press, and Karthik~R. Narasimhan.
\newblock Swe-bench: Can language models resolve real-world github issues?
\newblock In \emph{The Twelfth International Conference on Learning
  Representations, {ICLR} 2024, Vienna, Austria, May 7-11, 2024}.
  OpenReview.net, 2024.
\newblock URL \url{https://openreview.net/forum?id=VTF8yNQM66}.

\bibitem[Kaur and Singh(2015)]{kaur2015review}
Uttamjit Kaur and Gagandeep Singh.
\newblock A review on software maintenance issues and how to reduce maintenance
  efforts.
\newblock \emph{International Journal of Computer Applications}, 118\penalty0
  (1):\penalty0 6--11, 2015.

\bibitem[{Kimi Team}(2025{\natexlab{a}})]{kimi_k2_2025}
{Kimi Team}.
\newblock Kimi k2: Open agentic intelligence.
\newblock \emph{arXiv preprint arXiv:2507.20534}, 2025{\natexlab{a}}.

\bibitem[{Kimi Team}(2025{\natexlab{b}})]{team2025kimi}
{Kimi Team}.
\newblock Kimi k2: Open agentic intelligence.
\newblock \emph{arXiv preprint arXiv:2507.20534}, 2025{\natexlab{b}}.

\bibitem[{Kimi Team}(2026)]{kimi_k25_2026}
{Kimi Team}.
\newblock Kimi k2.5: Visual agentic intelligence.
\newblock \emph{arXiv preprint arXiv:2602.02276}, 2026.

\bibitem[Kuang et~al.(2025)Kuang, Li, Zhang, Li, Yin, Sun, Shen, and
  Yu]{kuang2025enconda}
Jiayi Kuang, Yinghui Li, Xin Zhang, Yangning Li, Di~Yin, Xing Sun, Ying Shen,
  and Philip~S. Yu.
\newblock Process-level trajectory evaluation for environment configuration in
  software engineering agents.
\newblock \emph{arXiv preprint arXiv:2510.25694}, 2025.

\bibitem[Li et~al.(2022)Li, Choi, Chung, Kushman, Schrittwieser, Leblond,
  Eccles, Keeling, Gimeno, Dal~Lago, et~al.]{li2022competition}
Yujia Li, David Choi, Junyoung Chung, Nate Kushman, Julian Schrittwieser,
  R{\'e}mi Leblond, Tom Eccles, James Keeling, Felix Gimeno, Agustin Dal~Lago,
  et~al.
\newblock Competition-level code generation with alphacode.
\newblock \emph{Science}, 378\penalty0 (6624):\penalty0 1092--1097, 2022.

\bibitem[Luo et~al.(2025)Luo, Jain, Singh, Tan, Patel, Wu, Ariyak, Cai, Venkat,
  Athiwaratkun, et~al.]{luodeepswe}
Michael Luo, Naman Jain, Jaskirat Singh, Sijun Tan, Ameen Patel, Qingyang Wu,
  Alpay Ariyak, Colin Cai, Shang Zhu~Tarun Venkat, Ben Athiwaratkun, et~al.
\newblock Deepswe: Training a fully open-sourced, state-of-the-art coding agent
  by scaling rl, 2025.

\bibitem[Manh et~al.(2023)Manh, Hai, Dau, Nguyen, Nghiem, Guo, and
  Bui]{manh2023vault}
Dung~Nguyen Manh, Nam~Le Hai, Anh~TV Dau, Anh~Minh Nguyen, Khanh Nghiem, Jin
  Guo, and Nghi~DQ Bui.
\newblock The vault: A comprehensive multilingual dataset for advancing code
  understanding and generation.
\newblock \emph{arXiv preprint arXiv:2305.06156}, 2023.

\bibitem[Mei et~al.(2025)Mei, Yao, Ge, Wang, Bi, Cai, Liu, Li, Li, Zhang,
  et~al.]{mei2025survey}
Lingrui Mei, Jiayu Yao, Yuyao Ge, Yiwei Wang, Baolong Bi, Yujun Cai, Jiazhi
  Liu, Mingyu Li, Zhong-Zhi Li, Duzhen Zhang, et~al.
\newblock A survey of context engineering for large language models.
\newblock \emph{arXiv preprint arXiv:2507.13334}, 2025.

\bibitem[Nguyen et~al.(2025{\natexlab{a}})Nguyen, Phan, Le~Hai, Doan, Nguyen,
  Pham, and Bui]{nguyencodemmlu}
Dung~Manh Nguyen, Thang~Chau Phan, Nam Le~Hai, Tien-Thong Doan, Nam~V Nguyen,
  Quang Pham, and Nghi~DQ Bui.
\newblock Codemmlu: A multi-task benchmark for assessing code understanding \&
  reasoning capabilities of codellms.
\newblock In \emph{The Thirteenth International Conference on Learning
  Representations}, 2025{\natexlab{a}}.

\bibitem[Nguyen et~al.(2025{\natexlab{b}})Nguyen, Chau, Nguyen, and
  Bui]{nguyen2025agilecoder}
Minh~Huynh Nguyen, Thang~Phan Chau, Phong~X Nguyen, and Nghi~DQ Bui.
\newblock Agilecoder: Dynamic collaborative agents for software development
  based on agile methodology.
\newblock In \emph{2025 IEEE/ACM Second International Conference on AI
  Foundation Models and Software Engineering (Forge)}, pages 156--167. IEEE,
  2025{\natexlab{b}}.

\bibitem[{OpenAI}(2024{\natexlab{a}})]{openai_gpt4o_2024}
{OpenAI}.
\newblock Hello gpt-4o.
\newblock \url{https://openai.com/index/hello-gpt-4o/}, 2024{\natexlab{a}}.
\newblock Accessed 2026-03-12.

\bibitem[{OpenAI}(2024{\natexlab{b}})]{sbv}
{OpenAI}.
\newblock Swe-bench verified, 2024{\natexlab{b}}.
\newblock \url{https://openai.com/index/introducing-swe-bench-verified/}.

\bibitem[{OpenAI}(2025{\natexlab{a}})]{openai_gpt41_2025}
{OpenAI}.
\newblock Introducing gpt-4.1 in the api.
\newblock \url{https://openai.com/index/gpt-4-1/}, 2025{\natexlab{a}}.
\newblock Accessed 2026-03-12.

\bibitem[{OpenAI}(2025{\natexlab{b}})]{openai_gpt5}
{OpenAI}.
\newblock Gpt-5 system card.
\newblock Technical report, OpenAI, 2025{\natexlab{b}}.
\newblock URL \url{https://cdn.openai.com/gpt-5-system-card.pdf}.

\bibitem[{OpenAI}(2025{\natexlab{c}})]{openai_gpt52_2025}
{OpenAI}.
\newblock Introducing gpt-5.2.
\newblock \url{https://openai.com/index/introducing-gpt-5-2/},
  2025{\natexlab{c}}.
\newblock Accessed 2026-03-12.

\bibitem[{OpenAI}(2025{\natexlab{d}})]{openai_gpt5_2025}
{OpenAI}.
\newblock Gpt-5 system card.
\newblock \url{https://openai.com/index/gpt-5-system-card/},
  2025{\natexlab{d}}.
\newblock Accessed 2026-03-12.

\bibitem[{OpenAI}(2025{\natexlab{e}})]{openai_gptoss_2025}
{OpenAI}.
\newblock gpt-oss-120b \& gpt-oss-20b model card.
\newblock
  \url{https://cdn.openai.com/pdf/419b6906-9da6-406c-a19d-1bb078ac7637/oai_gpt-oss_model_card.pdf},
  2025{\natexlab{e}}.
\newblock Accessed 2026-03-12.

\bibitem[{OpenAI}(2025{\natexlab{f}})]{openai_o3_2025}
{OpenAI}.
\newblock Introducing openai o3 and o4-mini.
\newblock \url{https://openai.com/index/introducing-o3-and-o4-mini/},
  2025{\natexlab{f}}.
\newblock Accessed 2026-03-12.

\bibitem[{OpenAI}(2026)]{openai_gpt54_2026}
{OpenAI}.
\newblock Introducing gpt-5.4.
\newblock \url{https://openai.com/index/introducing-gpt-5-4/}, 2026.
\newblock Accessed 2026-03-12.

\bibitem[Packer et~al.(2023)]{packer2023memgpt}
Charles Packer et~al.
\newblock Memgpt: Towards llms as operating systems.
\newblock \emph{arXiv preprint arXiv:2310.08560}, 2023.

\bibitem[Pan et~al.(2024)Pan, Wang, Neubig, Jaitly, Ji, Suhr, and
  Zhang]{pan2024training}
Jiayi Pan, Xingyao Wang, Graham Neubig, Navdeep Jaitly, Heng Ji, Alane Suhr,
  and Yizhe Zhang.
\newblock Training software engineering agents and verifiers with swe-gym,
  2024.
\newblock URL \url{https://arxiv.org/abs/2412.21139}.

\bibitem[Pham et~al.(2025)Pham, Phan, Phan, Chi, Nguyen, and
  Bui]{pham2025swesynth}
Minh~VT Pham, Huy~N Phan, Hoang~N Phan, Cuong~Le Chi, Tien~N Nguyen, and
  Nghi~DQ Bui.
\newblock Swe-synth: Synthesizing verifiable bug-fix data to enable large
  language models in resolving real-world bugs.
\newblock \emph{arXiv preprint arXiv:2504.14757}, 2025.

\bibitem[Phan et~al.(2024)Phan, Nguyen, and Bui]{phan2024hyperagent}
Huy~Nhat Phan, Phong~X. Nguyen, and Nghi D.~Q. Bui.
\newblock Hyperagent: Generalist software engineering agents to solve coding
  tasks at scale.
\newblock \emph{CoRR}, abs/2409.16299, 2024.
\newblock \doi{10.48550/ARXIV.2409.16299}.
\newblock URL \url{https://doi.org/10.48550/arXiv.2409.16299}.

\bibitem[Qian et~al.(2024)Qian, Han, Fung, Qin, Liu, and
  Ji]{qian2024creatortoolcreationdisentangling}
Cheng Qian, Chi Han, Yi~R. Fung, Yujia Qin, Zhiyuan Liu, and Heng Ji.
\newblock Creator: Tool creation for disentangling abstract and concrete
  reasoning of large language models, 2024.

\bibitem[Qiu et~al.(2025)Qiu, Qi, Zhang, Juan, Guo, Lu, Wang, Yao, Ren, Jiang,
  Zhou, Liu, Yang, Wu, Huang, Liu, Wang, and
  Wang]{qiu2025alitageneralistagentenabling}
Jiahao Qiu, Xuan Qi, Tongcheng Zhang, Xinzhe Juan, Jiacheng Guo, Yifu Lu, Yimin
  Wang, Zixin Yao, Qihan Ren, Xun Jiang, Xing Zhou, Dongrui Liu, Ling Yang, Yue
  Wu, Kaixuan Huang, Shilong Liu, Hongru Wang, and Mengdi Wang.
\newblock Alita: Generalist agent enabling scalable agentic reasoning with
  minimal predefinition and maximal self-evolution, 2025.

\bibitem[{Qwen Team}(2025)]{qwen3coder_2025}
{Qwen Team}.
\newblock Qwen3-coder.
\newblock \url{https://github.com/QwenLM/Qwen3-Coder}, 2025.
\newblock Accessed 2026-03-12.

\bibitem[Robeyns et~al.(2025)Robeyns, Szummer, and Aitchison]{robeyns2025self}
Maxime Robeyns, Martin Szummer, and Laurence Aitchison.
\newblock A self-improving coding agent.
\newblock \emph{arXiv preprint arXiv:2504.15228}, 2025.

\bibitem[Sarthi et~al.(2024)Sarthi, Abdullah, Tuli, Khanna, Goldie, and
  Manning]{sarthi2024raptor}
Parth Sarthi, Salman Abdullah, Aditi Tuli, Shubh Khanna, Anna Goldie, and
  Christopher~D. Manning.
\newblock Raptor: Recursive abstractive processing for tree-organized
  retrieval.
\newblock \emph{International Conference on Learning Representations}, 2024.

\bibitem[{Scale AI}(2025)]{deng2025swe}
{Scale AI}.
\newblock Swe-bench pro: Can ai agents solve long-horizon software engineering
  tasks?
\newblock \emph{arXiv preprint arXiv:2509.16941}, 2025.

\bibitem[Singh et~al.(2019)Singh, Sharma, and Kumar]{singh2019analysis}
Chamkaur Singh, Neeraj Sharma, and Narender Kumar.
\newblock Analysis of software maintenance cost affecting factors and
  estimation models.
\newblock \emph{Int. J. Sci. Technol. Res}, 8\penalty0 (9):\penalty0 276--281,
  2019.

\bibitem[To et~al.(2023)To, Nguyen, and Bui]{to2023functional}
Hung~Quoc To, Minh~Huynh Nguyen, and Nghi~DQ Bui.
\newblock Functional overlap reranking for neural code generation.
\newblock \emph{arXiv preprint arXiv:2311.03366}, 2023.

\bibitem[Wang et~al.(2024{\natexlab{a}})Wang, Xie, Jiang, Mandlekar, Xiao, Zhu,
  Fan, and Anandkumar]{wang2024voyager}
Guanzhi Wang, Yuqi Xie, Yunfan Jiang, Ajay Mandlekar, Chaowei Xiao, Yuke Zhu,
  Linxi Fan, and Anima Anandkumar.
\newblock Voyager: An open-ended embodied agent with large language models.
\newblock \emph{Transactions on Machine Learning Research}, 2024{\natexlab{a}}.

\bibitem[Wang et~al.(2024{\natexlab{b}})Wang, Yang, Wang, Huang, Chu, Song,
  Zhang, Chen, and Ma]{wang2024testeval}
Wenhan Wang, Chenyuan Yang, Zhijie Wang, Yuheng Huang, Zhaoyang Chu, Da~Song,
  Lingming Zhang, An~Ran Chen, and Lei Ma.
\newblock Testeval: Benchmarking large language models for test case
  generation.
\newblock \emph{arXiv preprint arXiv:2406.04531}, 2024{\natexlab{b}}.

\bibitem[Wang et~al.(2025)Wang, Pi{\k{e}}kos, Nanbo, Laakom, Chen, Ostaszewski,
  Zhuge, and Schmidhuber]{wang2025huxley}
Wenyi Wang, Piotr Pi{\k{e}}kos, Li~Nanbo, Firas Laakom, Yimeng Chen, Mateusz
  Ostaszewski, Mingchen Zhuge, and J{\"u}rgen Schmidhuber.
\newblock Huxley-g{\"o}del machine: Human-level coding agent development by an
  approximation of the optimal self-improving machine, 2025.

\bibitem[Wang et~al.(2024{\natexlab{c}})Wang, Chen, Yuan, Zhang, Li, Peng, and
  Ji]{wang2024executable}
Xingyao Wang, Yangyi Chen, Lifan Yuan, Yizhe Zhang, Yunzhu Li, Hao Peng, and
  Heng Ji.
\newblock Executable code actions elicit better llm agents.
\newblock In \emph{Forty-first International Conference on Machine Learning},
  2024{\natexlab{c}}.

\bibitem[Wang et~al.(2024{\natexlab{d}})Wang, Li, Song, Xu, Tang, Zhuge, Pan,
  Song, Li, Singh, et~al.]{wang2024openhands}
Xingyao Wang, Boxuan Li, Yufan Song, Frank~F Xu, Xiangru Tang, Mingchen Zhuge,
  Jiayi Pan, Yueqi Song, Bowen Li, Jaskirat Singh, et~al.
\newblock Openhands: An open platform for ai software developers as generalist
  agents.
\newblock \emph{arXiv preprint arXiv:2407.16741}, 2024{\natexlab{d}}.

\bibitem[Wang et~al.(2023)Wang, Le, Gotmare, Bui, Li, and Hoi]{wang2023codet5+}
Yue Wang, Hung Le, Akhilesh~Deepak Gotmare, Nghi~DQ Bui, Junnan Li, and
  Steven~CH Hoi.
\newblock Codet5+: Open code large language models for code understanding and
  generation.
\newblock \emph{arXiv preprint arXiv:2305.07922}, 2023.

\bibitem[Wang et~al.(2024{\natexlab{e}})Wang, Fried, and
  Neubig]{wang2024troveinducingverifiableefficient}
Zhiruo Wang, Daniel Fried, and Graham Neubig.
\newblock Trove: Inducing verifiable and efficient toolboxes for solving
  programmatic tasks, 2024{\natexlab{e}}.

\bibitem[Wei et~al.(2023)Wei, Wang, Liu, Ding, and Zhang]{wei2023magicoder}
Yuxiang Wei, Zhe Wang, Jiawei Liu, Yifeng Ding, and Lingming Zhang.
\newblock Magicoder: Source code is all you need.
\newblock \emph{arXiv preprint arXiv:2312.02120}, 2023.

\bibitem[Wei et~al.(2025)Wei, Duchenne, Copet, Carbonneaux, Zhang, Fried,
  Synnaeve, Singh, and Wang]{wei2025swerl}
Yuxiang Wei, Olivier Duchenne, Jade Copet, Quentin Carbonneaux, Lingming Zhang,
  Daniel Fried, Gabriel Synnaeve, Rishabh Singh, and Sida~I. Wang.
\newblock Swe-rl: Advancing llm reasoning via reinforcement learning on open
  software evolution.
\newblock \emph{arXiv preprint arXiv:2502.18449}, 2025.

\bibitem[Xia and Zhang(2023)]{xia2023conversational}
Chunqiu~Steven Xia and Lingming Zhang.
\newblock Conversational automated program repair.
\newblock \emph{arXiv preprint arXiv:2301.13246}, 2023.

\bibitem[Xia and Zhang(2024)]{xia2024automated}
Chunqiu~Steven Xia and Lingming Zhang.
\newblock Automated program repair via conversation: Fixing 162 out of 337 bugs
  for \$0.42 each using chatgpt.
\newblock In \emph{Proceedings of the 33rd ACM SIGSOFT International Symposium
  on Software Testing and Analysis}, pages 819--831, 2024.

\bibitem[Xia et~al.(2024)Xia, Deng, Dunn, and Zhang]{xia2024agentless}
Chunqiu~Steven Xia, Yinlin Deng, Soren Dunn, and Lingming Zhang.
\newblock Agentless: Demystifying llm-based software engineering agents.
\newblock \emph{arXiv preprint arXiv:2407.01489}, 2024.

\bibitem[Yang et~al.(2025)Yang, Li, Yang, Zhang, Hui, Zheng, Yu, Gao, Huang,
  Lv, et~al.]{qwen3_2025}
An~Yang, Anfeng Li, Baosong Yang, Beichen Zhang, Binyuan Hui, Bo~Zheng, Bowen
  Yu, Chang Gao, Chengen Huang, Chenxu Lv, et~al.
\newblock Qwen3 technical report.
\newblock \emph{arXiv preprint arXiv:2505.09388}, 2025.

\bibitem[Yang et~al.(2024{\natexlab{a}})Yang, Jimenez, Wettig, Lieret, Yao,
  Narasimhan, and Press]{yang2024swe}
John Yang, Carlos~E Jimenez, Alexander Wettig, Kilian Lieret, Shunyu Yao,
  Karthik Narasimhan, and Ofir Press.
\newblock Swe-agent: Agent-computer interfaces enable automated software
  engineering.
\newblock \emph{arXiv preprint arXiv:2405.15793}, 2024{\natexlab{a}}.

\bibitem[Yang et~al.(2024{\natexlab{b}})Yang, Jimenez, Zhang, Lieret, Yang, Wu,
  Press, Muennighoff, Synnaeve, Narasimhan, et~al.]{yang2024swe2}
John Yang, Carlos~E Jimenez, Alex~L Zhang, Kilian Lieret, Joyce Yang, Xindi Wu,
  Ori Press, Niklas Muennighoff, Gabriel Synnaeve, Karthik~R Narasimhan, et~al.
\newblock Swe-bench multimodal: Do ai systems generalize to visual software
  domains?
\newblock \emph{arXiv preprint arXiv:2410.03859}, 2024{\natexlab{b}}.

\bibitem[Ye et~al.(2026)Ye, He, Arak, Dong, and Song]{ye2026meta}
Haoran Ye, Xuning He, Vincent Arak, Haonan Dong, and Guojie Song.
\newblock Meta context engineering via agentic skill evolution.
\newblock \emph{arXiv preprint arXiv:2601.21557}, 2026.

\bibitem[Yuan et~al.(2024)Yuan, Liu, Ding, Wang, Chen, Peng, and
  Lou]{yuan2024evaluating}
Zhiqiang Yuan, Mingwei Liu, Shiji Ding, Kaixin Wang, Yixuan Chen, Xin Peng, and
  Yiling Lou.
\newblock Evaluating and improving chatgpt for unit test generation.
\newblock \emph{Proceedings of the ACM on Software Engineering}, 1\penalty0
  (FSE):\penalty0 1703--1726, 2024.

\bibitem[{Z.ai}(2025)]{zai_glm47_2025}
{Z.ai}.
\newblock Glm-4.7: Advancing the coding capability.
\newblock \url{https://z.ai/blog/glm-4.7}, 2025.
\newblock Accessed 2026-03-12.

\bibitem[{Z.ai Team}(2026)]{zai_glm5_2026}
{Z.ai Team}.
\newblock Glm-5: From vibe coding to agentic engineering.
\newblock \emph{arXiv preprint arXiv:2602.15763}, 2026.

\bibitem[Zan et~al.(2025)Zan, Huang, Liu, Chen, Zhang, Xin, Chen, Liu, Zhong,
  Li, et~al.]{zan2025multi}
Daoguang Zan, Zhirong Huang, Wei Liu, Hanwu Chen, Linhao Zhang, Shulin Xin,
  Lu~Chen, Qi~Liu, Xiaojian Zhong, Aoyan Li, et~al.
\newblock Multi-swe-bench: A multilingual benchmark for issue resolving.
\newblock \emph{arXiv preprint arXiv:2504.02605}, 2025.

\bibitem[Zhang et~al.(2025{\natexlab{a}})Zhang, Hu, Lu, Lange, and
  Clune]{zhang2025darwin}
Jenny Zhang, Shengran Hu, Cong Lu, Robert Lange, and Jeff Clune.
\newblock Darwin godel machine: Open-ended evolution of self-improving agents.
\newblock \emph{arXiv preprint arXiv:2505.22954}, 2025{\natexlab{a}}.

\bibitem[Zhang et~al.(2025{\natexlab{b}})Zhang, He, Zhang, Kang, Li, Xie, Wang,
  Wang, Huang, Fu, Nallipogu, Lin, Dang, Rajmohan, and Zhang]{zhang2025swe}
Linghao Zhang, Shilin He, Chaoyun Zhang, Yu~Kang, Bowen Li, Chengxing Xie,
  Junhao Wang, Maoquan Wang, Yufan Huang, Shengyu Fu, Elsie Nallipogu, Qingwei
  Lin, Yingnong Dang, Saravan Rajmohan, and Dongmei Zhang.
\newblock Swe-bench goes live!
\newblock \emph{arXiv preprint arXiv:2505.23419}, 2025{\natexlab{b}}.

\bibitem[Zhang et~al.(2023)Zhang, Fang, Xie, Zhang, Yang, Sun, Yu, and
  Chen]{zhang2023survey}
Quanjun Zhang, Chunrong Fang, Yang Xie, Yaxin Zhang, Yun Yang, Weisong Sun,
  Shengcheng Yu, and Zhenyu Chen.
\newblock A survey on large language models for software engineering.
\newblock \emph{arXiv preprint arXiv:2312.15223}, 2023.

\bibitem[Zhang et~al.(2024)Zhang, Ruan, Fan, and
  Roychoudhury]{zhang2024autocoderover}
Yuntong Zhang, Haifeng Ruan, Zhiyu Fan, and Abhik Roychoudhury.
\newblock Autocoderover: Autonomous program improvement.
\newblock In \emph{Proceedings of the 33rd ACM SIGSOFT International Symposium
  on Software Testing and Analysis}, pages 1592--1604, Vienna, Austria, 2024.
  ACM.

\bibitem[Zhong et~al.(2024)Zhong, Guo, Gao, Ye, and Wang]{zhong2023memorybank}
Wanjun Zhong, Lianghong Guo, Qiqi Gao, He~Ye, and Yanlin Wang.
\newblock Memorybank: Enhancing large language models with long-term memory.
\newblock \emph{AAAI Conference on Artificial Intelligence}, 2024.

\bibitem[Zhou et~al.(2023)Zhou, Yan, Shlapentokh-Rothman, Wang, and
  Wang]{zhou2023language}
Andy Zhou, Kai Yan, Michal Shlapentokh-Rothman, Haohan Wang, and Yu-Xiong Wang.
\newblock Language agent tree search unifies reasoning acting and planning in
  language models.
\newblock \emph{arXiv preprint arXiv:2310.04406}, 2023.

\bibitem[Zhuo et~al.(2024)Zhuo, Vu, Chim, Hu, Yu, Widyasari, Yusuf, Zhan, He,
  Paul, et~al.]{zhuo2024bigcodebench}
Terry~Yue Zhuo, Minh~Chien Vu, Jenny Chim, Han Hu, Wenhao Yu, Ratnadira
  Widyasari, Imam Nur~Bani Yusuf, Haolan Zhan, Junda He, Indraneil Paul, et~al.
\newblock Bigcodebench: Benchmarking code generation with diverse function
  calls and complex instructions.
\newblock \emph{arXiv preprint arXiv:2406.15877}, 2024.

\end{thebibliography}
